\documentclass[preprint, 11pt,authoryear]{elsarticle}
\usepackage{amsmath}
\usepackage{graphicx,psfrag,epsf}
\usepackage{enumerate}
\usepackage{enumitem}
\usepackage{adjustbox}
\usepackage{natbib}
\usepackage{url} 

\DeclareMathOperator*{\argmax}{arg\,max}
\DeclareMathOperator*{\argmin}{arg\,min}
\usepackage{amsfonts}
\usepackage{verbatim}
\usepackage{xcolor}
\usepackage{amssymb}
\usepackage{epstopdf}%
\usepackage{caption,subcaption}
\usepackage{algorithm,algorithmicx}
\usepackage[noend]{algpseudocode}
\usepackage[detect-all]{siunitx}
\usepackage{tabu}
\usepackage{arydshln}
\usepackage{eufrak}
\usepackage{mathtools}
\usepackage{hanging}

\newcolumntype{"}{@{\hskip\tabcolsep\vrule width 1pt\hskip\tabcolsep}}
\makeatother
\providecommand{\U}[1]{\protect\rule{.1in}{.1in}}

\newcommand{\bbet}{\boldsymbol{\beta}}
\newcommand{\tbbet}{\boldsymbol{\tilde{\beta}}}
\newcommand{\bbbet}{\boldsymbol{\bar{\beta}}}
\newcommand{\hbbet}{\boldsymbol{\hat{\beta}}}

\newcommand{\bx}{\mathbf{x}}
\newcommand{\bX}{\mathbf{X}}
\newcommand{\by}{\mathbf{y}}

\algrenewcommand\algorithmicrequire{\hspace*{\algorithmicindent}\textbf{Input:}}
\algrenewcommand{\algorithmicensure}{\hspace*{\algorithmicindent}\textbf{Initialize:}}
\algdef{SE}[DOWHILE]{Do}{doWhile}{\algorithmicdo}[1]{\algorithmicwhile\ #1}%

\usepackage{placeins}
\usepackage{booktabs}
\usepackage{multirow}
\usepackage{rotating}
\usepackage{afterpage}
\usepackage{setspace}
\usepackage{mathtools}
\usepackage{hyperref}
\hypersetup{
	colorlinks=true,
	linkcolor=blue,
	citecolor=blue,
	filecolor=magenta,
	linktocpage=true,      
	urlcolor=blue,
}

\begin{document}
	
	\begin{frontmatter}
		
		\title{\textbf{Multi-Model Subset Selection}}
		
		\author[ubc]{Anthony-Alexander Christidis\corref{cor1}}
		\ead{anthony.christidis@stat.ubc.ca}
		
		\author[kuleuven]{Stefan Van Aelst}
		\ead{stefan.vanaelst@kuleuven.be}
		
		\author[ubc]{Ruben Zamar}
		\ead{ruben@stat.ubc.ca}
		
		\affiliation[ubc]{%
			organization={Department of Statistics, University of British Columbia},%
			addressline={2207 Main Mall}, 
			city={Vancouver},
			postcode={BC V6T 1Z4},
			country={Canada}
		}
		
		\affiliation[kuleuven]{%
			organization={Department of Mathematics, KU Leuven},%
			addressline={Celestijnenlaan 200B}, 
			city={Leuven},
			postcode={3001},
			country={Belgium}
		}
		
		\cortext[cor1]{Corresponding author. Tel: +1-514-945-4237; Email: \texttt{anthony.christidis@stat.ubc.ca}.}

		\begin{abstract}
			The two primary approaches for high-dimensional regression problems are sparse methods (e.g., best subset selection, which uses the $ \ell_0 $-norm in the penalty) and ensemble methods (e.g., random forests). Although sparse methods typically yield interpretable models, in terms of prediction accuracy they are often outperformed by "blackbox" multi-model ensemble methods. A regression ensemble is introduced which combines the interpretability of sparse methods with the high prediction accuracy of ensemble methods.
			An algorithm is proposed to solve the joint optimization of the corresponding $ \ell_0 $-penalized regression models by extending recent developments in $ \ell_0 $-optimization for sparse methods to multi-model regression ensembles. The sparse and diverse models in the ensemble are learned simultaneously from the data. Each of these models provides an explanation for the relationship between a subset of predictors and the response variable. Empirical studies and theoretical knowledge about ensembles are used to gain insight into the ensemble method's performance, focusing on the interplay between bias, variance, covariance, and variable selection. In prediction tasks, the ensembles can outperform state-of-the-art competitors on both simulated and real data. Forward stepwise regression is also generalized to multi-model regression ensembles and used to obtain an initial solution for the algorithm. The optimization algorithms are implemented in publicly available software packages.
		\end{abstract}
		
		
	\end{frontmatter}
	
	\section{Introduction}
	
	\onehalfspacing
	In modern machine learning tasks, it is often not sufficient to create a model with good prediction accuracy. In high-stakes data-driven decision making it is often necessary to obtain a model that is also highly interpretable so that predictions are explainable. Recently, there have appeared several influential articles advocating for the need of statistical and machine learning procedures that retain a certain degree of interpretability on top of their high prediction accuracy, see e.g. relevant discussions in \cite{rudin2019stop} and \cite{murdoch2019definitions}. The issue of interpretability is particularly important when analyzing data where the number of predictors $ p $ is much larger than the number of samples $ n $ ($ p \gg n $). For such high-dimensional data,  parsimonious models where only a small subset 
	of the predictors are included are often preferred. For example, with the development of new genomic data collecting and processing  technologies, the expression levels for thousands of genes are stored in computer systems for further analysis. A valuable model would be a model with high prediction accuracy which at the same time identifies only a small subset of genes as relevant to predict the outcome of interest. 
	
	To address the problem of prediction accuracy and interpretability for high-dimensional data, sparse  methods have been developed over the last decades. In essence, sparse  methods optimize the goodness-of-fit of a single model while restricting or penalizing its complexity, resulting in an interpretable model with good prediction accuracy. Sparse regularization methods have been developed for a variety of model classes, see e.g. \cite{hastie2019statistical} for an extensive treatment.
	While prediction is an important aspect of sparse  methods,  there is also a strong emphasis on interpretability by aiming to uncover the relation between the output and candidate predictors. A thorough theoretical treatment of sparse regularization methods and variable selection can be found in \cite{buhlmann2011statistics}. 
	
	While sparse  methods have well-established statistical theory and result in interpretable models, they are often outperformed by ensemble methods in terms of prediction accuracy. Ensemble methods, which generate and aggregate multiple diverse models, are among the most popular ``blackbox" algorithms for the analysis of high-dimensional data. They have led to a plethora of successful applications in fraud detection \citep[see e.g.][]{kim2012stock, louzada2012bagging}, computer vision \citep[see e.g.][]{
		yu2015image}, genetics \citep[see e.g.][]{dorani2018ensemble, genes11070819}, speech recognition \citep[see e.g.][]{
		rieger2014speech},  and many other fields. Diversity among the aggregated models is essential for the good predictive performance of ensembles \citep{brown2005managing}. Current state-of-the-art methods usually rely on randomization or the sequential refitting of residuals to achieve diversity, which results in ensembles comprised of a large number of uninterpretable models with poor individual prediction accuracy.
	
	We aim to combine the interpretability of sparse  methods with the high prediction accuracy of ensemble methods. To this end, we leverage recent developments in $ \ell_0 $-constrained optimization, and generalize the algorithm to multi-model regression ensembles. The proposed methodology yields ensembles consisting of a small number of sparse and diverse models, learned jointly from the data, which each have a high prediction accuracy. Hence, each model in the ensemble provides a  possible explanation
	for the relationship between the predictors and the response. By using a cross-validation (CV) criterion the degree to which the models are sparse and diverse is also driven directly by the data. 
	
	The newly proposed ensembles outperform state-of-the-art sparse and ensemble methods, in terms of prediction performance and variable selection (recall and precision), in an extensive simulation study that includes a large number of high-dimensional scenarios. 
	We investigate the good performance of our method, and show that the proposed ensembles learn the optimal balance between individual model accuracy and diversity between them.
	In  a  gene expression data application our method generated ensembles comprised of individual models with prediction accuracy on par with models generated by sparse methods. Moreover, the ensembling of a small number of these highly accurate individual models was competitive with ``blackbox" ensemble methods utilizing a large number of weak uninterpretable models. We also show how the proposed ensembles can be used to rank genes in order of importance. 
	
	The methodology introduced in this article is related to the work of \cite{christidis2020split} who generalized sparse regularization methods to multi-model regression ensembles. They proposed to optimize a global objective function that penalizes the complexity of the models in the ensemble and the overlap between the models to encourage diversity between them. Their method can be seen as a multi-convex relaxation of our proposal, which we bypass by optimizing the ensembles directly. For completeness, we compare their approach to our proposal in this article, and include their method as a competitor in our numerical experiments.
	
	The remainder of this article is organized as follows. In Section \ref{sec:literature} we shortly review both sparse and ensemble methods. In Section \ref{sec:BSpS} we introduce a  framework unifying sparse and ensemble methods which forms the basis of our methodology, and  review related work in the literature. In Section \ref{sec:stepwise}, we generalize stepwise regression to multi-model ensembles, which then constitutes the initial solution for the  multi-model ensembling algorithm of Section \ref{sec:PSGD}. This projected subsets gradient descent algorithm adapts $ \ell_0 $ optimization approaches to multi-model regression ensembles. In Section \ref{sec:simulation} we perform an extensive simulation study to compare the proposed methodology to state-of-the-art methods. 
	In Section \ref{sec:empirical_comparison} we combine empirical studies with theoretical knowledge about ensembles to gain insight into our ensemble method's performance, focusing on the interplay between bias, variance, and covariance, and their connection to model accuracy and diversity.
	In Section \ref{sec:eye} we illustrate the good performance of the proposed methodology on real data and show how the proposed ensembles can be used to rank predictors in order of importance. Section \ref{sec:discussion} closes the article with a discussion.
	
	\section{Sparse and Ensemble Methods} \label{sec:literature}
	
	We consider a dataset consisting of a response vector  $\mathbf{y}=(y_{1},\dots, y_{n})^T \in \mathbb{R}^n $  and a design matrix $ \bX \in \mathbb{R}^{n \times p} $ containing $ n $ observations $ \bx_i=(x_{i1},\dots,x_{ip})^T\ (i=1,\dots,n)$ for $ p $ predictors and assume the linear model
	\begin{equation*}
		y_{i} = \mathbf{x}_{i}^{\prime} \boldsymbol{\beta}_{0} + \sigma \epsilon_{i},
		\quad 1\leq i \leq n,
	\end{equation*}
	where  $ \bbet_{0} \in \mathbb{R}^p$ and the elements of the noise vector $\boldsymbol{\epsilon} = (\epsilon_1, \dots, \epsilon_n)^T \in \mathbb{R}^n$ are independent and identically distributed with variance 1. We focus on the high-dimensional setting, i.e. $ p \gg n $ where the underlying model is sparse, meaning that only a small fraction of the available predictors are relevant for explaining the response.
	We assume that the response $ \by $ and the columns of the design matrix are standardized, that is 
	\begin{align*}
		\frac{1}{n}\sum\limits_{i=1}^{n}x_{ij}=0,\quad 
		\frac{1}{n}\sum\limits_{i=1}^{n}x_{ij}^{2}=1,\quad 1\leq j \leq p, \quad 
		\frac{1}{n}\sum\limits_{i=1}^{n}y_{i}=0 \quad \text{and} \quad
		\frac{1}{n}\sum\limits_{i=1}^{n}y_{i}^{2}=1,
	\end{align*}
	so that we can omit the intercept term from the regression model.
	
	\subsection{Sparse Methods}
	
	Sparse regularization methods favor sparse solutions by penalizing model complexity. The purpose of such methods is to find a sparse model that achieves good prediction accuracy. 	To explain the intuition for the potential reduction in prediction error of sparse regularization methods, consider an estimator $\hat{f}(\bx)=\hbbet^T\bx$ of the regression function $ f(\bx)= \bbet_0^T \bx $. The  mean squared prediction error (MSPE) of $\hat{f}$ may be decomposed into its bias, variance and irreducible error,
	\begin{align} \label{eq:MSPE}
		\text{MSPE}\left[\hat{f}\right] = \mathbb{E}_{\bx}\left[(f(\bx)-\hat{f}(\bx))^2\right] + \sigma^2 = \text{Bias}\left[\hat{f}\right]^2 + \text{Var}\left[\hat{f}\right] + \sigma^2.
	\end{align}
	Since least squares regression is the best linear unbiased estimator (BLUE), the rationale for regularized estimation is to exploit the bias-variance tradeoff favorably, i.e. to induce a small increase in bias in exchange for a larger decrease in variance.

	The most natural approach for sparse modeling is Best Subset Selection (BSS), first mentioned in the literature by \cite{garside1965best}, which solves the nonconvex problem
	\begin{align} \label{BSS}
		\min_{\bbet \in \mathbb{R}^p} \lVert \by -\bX \bbet \rVert_2^2 \quad \text{subject to} \quad  \lVert \bbet \rVert_0 \leq t.
	\end{align}
	The number of nonzero coefficients $ t \leq \min(n-1, p) $ in the coefficient vector $ \bbet = (\beta_1, \beta_2, \dots, \beta_p)^T $ is typically determined in a data-driven way, e.g. by CV. While BSS has been shown to have desirable variable selection and estimation properties \citep[see e.g.][]{bunea2007aggregation, shen2013constrained}, it is an NP-hard problem \citep{welch1982algorithmic}. Indeed, the number of possible subsets that must be evaluated to determine the exact solution is given by
	\begin{align}
		\mathcal{K}(p, t) = \sum_{j=0}^{t} \binom{p}{j}.
	\end{align}
	For example, $\mathcal{K}(15, 10) = $ 30,826 which is already a large number of subsets for a setting with a small number of predictor variables. While many proposals have been made to determine the optimal subset based on the training data \citep[see e.g.][]{mallows1973some,akaike1974new,schwarz1978estimating}, CV is often recommended \citep{hastie2009elements} which makes the procedure even more computationally intensive.
	
	The branch-and-bound algorithm \citep{furnival1974regressions,gatu2006branch} was initially the procedure of choice for BSS, but this algorithm does not scale well beyond $ p=30 $. 
	To address this lack of computational feasibility, stepwise algorithms have been developed, see e.g. \cite{
		bendel1977comparison}, but 
	their popularity has decreased greatly in the last few decades due
	to questionable model selection properties, see e.g. \cite{
		roecker1991prediction}.
	
	To overcome the shortcomings of stepwise procedures, sparse regularization methods were popularized, first by basis pursuit denoising \citep{chen1994basis} but quickly followed by the closely related Lasso \citep{tibshirani1996regression}, a convex relaxation of BSS which solves problems of the form
	\begin{align} \label{eq:BSS}
		\min_{\bbet \in \mathbb{R}^p} \lVert \by -\bX \bbet \rVert_2^2 \quad \text{subject to} \quad  \lVert \bbet \rVert_1 \leq t.
	\end{align}
	Efficient convex solvers have been developed for the Lasso, see e.g. \cite{glmnet}. However, restrictive conditions on the covariance of the predictors must hold for the Lasso to have good variable selection properties \citep[see e.g.][]{zhao2006model} and good relative prediction error compared to BSS \citep[see e.g.][]{zhang2014lower}.
	To address these shortcomings, many other sparse regularization methods have  been proposed, see e.g. \cite{SCAD, zou2005regularization, zou2006adaptive, meinshausen2007relaxed,candes2007dantzig, zhang2010nearly}.

	\cite{bertsimas2016best} studied the nonconvex optimization problem \eqref{BSS} for BSS with a modern optimization lens which has led to new optimization strategies for $ \ell_0 $-penalized procedures, see e.g. \cite{bertsimas2020sparse, takano2020best, kenney2021mip}.
	\cite{bertsimas2016best} apply a projected gradient descent algorithm to generate good local solutions and then use these as warm-starts for Mixed Integer Optimization (MIO).
	\cite{thompson2022robust} adapted their algorithm to develop a robust version of BSS. This approach scales to problems of dimension $ p > $ 1,000, but even with the warm-starts this algorithm may still require over 30 minutes to compute the solution. To reduce the need for a MIO solver, \cite{hazimeh2020fast} proposed an alternative method. Once a local (incumbent) solution has been obtained from a projected gradient descent algorithm, they make small perturbations of this solution and apply the projected gradient descent algorithm to each of these perturbations. If the best solution obtained in this way improves the incumbent solution, then it is set as the new solution. This process is repeated until  the objective function does not decrease anymore. \cite{hazimeh2020fast} provide empirical evidence that their proposal often recovers either the optimal or a near-optimal solution to BSS in a matter of seconds, even for $ p> $ 1,000.

	\subsection{Ensemble Methods} \label{sec:ensembles}
	
	The advantage of ensemble methods in terms of prediction accuracy can also be seen from a decomposition of their MSPE. The MSPE of an ensemble $\bar{f} = \sum_{g=1}^G \hat{f}_g/G$ of $ G $ regression functions can be decomposed as
	\begin{align} 
		\text{MSPE}\left[\bar{f}\right] = \text{Bias}\left[\bar{f}\right]^2 + \text{Var}\left[\bar{f}\right] + \sigma^2, \label{eq:MSPE_ensemble}
	\end{align}
	with
	\begin{align} 
		\text{Bias}\left[\bar{f}\right] = \overline{\text{Bias}} 
		\quad \text{and} \quad		
		\text{Var}\left[\bar{f}\right] = \frac{1}{G} \, \overline{\text{Var}} + \frac{G-1}{G} \, \overline{\text{Cov}}, \label{eq:MSPE_variance}
	\end{align}
	where $ \overline{\text{Bias}} $, $ \overline{\text{Var}} $ and $ \overline{\text{Cov}} $ are the average biases, variances and pairwise covariances of the $ G $ regression functions in the ensemble \citep{ueda1996generalization}. From~\eqref{eq:MSPE_variance} it is clear that an ensemble can successfully reduce its variance if the models in the ensemble are 
	sufficiently diverse (uncorrelated in this regression context), especially if the number of models is large.
	
	Over the last twenty years the statistics and machine learning communities have  seen an increase in algorithmic approaches to generate ensembles, with most proposals relying on randomization \citep[see e.g.][]{RF, random_glm_paper} or boosting \cite[see e.g.][]{GBM, buhlmann2003boosting, boosting, chen2016xgboost}. These ensemble methods aim to generate a collection of diverse models, often  based on substantially different subsets of predictors. For example, in Random Forests random sampling of the data (bagging) \citep{breiman1996bagging} and the random predictor subspace method \citep{ho1998random} are combined to generate uncorrelated trees to achieve a lower generalization error \citep{RF}. 
	In gradient boosting, diverse members (typically decision trees) are generated by sequentially fitting the residuals of the previous fit. 

	Interpretability of such ensembles is typically unfeasible. However, several ad hoc methods have been developed to assess predictor importance \citep[see e.g.][]{hastie2009elements}. In an attempt to bridge the gap between interpretability and ensemble methods, \cite{buhlmann2006sparse} introduced sparse boosting by minimizing a penalized $\ell_2$-loss function for better variable selection.

	\section{Best Split Selection} \label{sec:BSpS}
	
	The multiplicity of good models is a phenomenon that has long been acknowledged, see e.g. relevant discussions in \cite{mccullagh1989monographs} and \cite{mountain1989combined}. Different, yet equally good models can provide distinct explanations for the underlying relationship between predictors and response. However,  current state-of-the-art ensemble methods lack interpretability due to the large number of models that is aggregated.
	Moreover, the individual models in the ensembles do not have high prediction accuracy on their own, but only work well when they are pooled together in the final ensemble fit. Each individual model is thus not insightful or reliable on its own.
	Hence, there currently is a gap between interpretable single model methods such as sparse regression methods and algorithmic ensemble methods.   
	We aim to fill this gap by developing a systematic approach to construct ensembles consisting of a relatively small number of interpretable sparse models with high individual prediction accuracy. Each of these models is learned directly from the data and provides a reliable relationship between the predictors and the response. Diversity between the models is imposed by restricting the sharing of predictors between different models, so that they are well-suited to be combined in an ensemble.

	Our framework that unifies sparse and ensemble methods aims to find a collection of $ G \geq 2$ sparse and diverse models that will be combined in an ensemble. Denote the matrix of model coefficients by
	\begin{align}
		\bbet_{1:G} = \begin{pmatrix}
			\beta_{1}^1 & \beta_{1}^2 & \dots & \beta_{1}^G \\
			\beta_{2}^1 & \beta_{2}^2 & \dots & \beta_{2}^G \\
			\vdots & \vdots & \ddots & \vdots \\
			\beta_{p}^1 & \beta_{p}^2 & \dots & \beta_{p}^G
		\end{pmatrix},
	\end{align}
	where $ \bbet_{1:G} \in \mathbb{R}^{p \times G} $ and $ \beta_j^g $ is the coefficient for predictor $ j $ in model $ g $, $ 1 \leq g \leq G $. For notational convenience let $ \bbet^g = (\beta_1^g, \beta_2^g, \dots, \beta_p^g)^T \in \mathbb{R}^p $ be the coefficients of model $ g $ and $ \bbet_{j\cdot} = (\beta_j^1, \beta_j^2, \dots, \beta_j^G)^T \in \mathbb{R}^G$ the coefficients of predictor $ j $ across the $ G $ models. 
	
	Best Split Selection (BSpS) aims to find $G$ sparse models, in such a way that each model explains well the response while at the same time the different models do not have much overlap. In this way, the models complement each other well  in an ensemble.  To reach this goal BSpS solves, for a fixed number of sparse models $ G $, the nonconvex problem
	\begin{align} \label{eq:BSpS}
		\min_{\bbet^1, \dots, \, \bbet^G \in \mathbb{R}^p} \sum_{g=1}^{G} \| \by - \bX \bbet^g \|_2^2 \quad \text{subject to} \quad \begin{cases}
			\lVert\bbet^g\rVert_0 \leq t, \, &1 \leq g \leq G, \\
			\lVert\bbet_{j\cdot}\rVert_0 \leq u, \, & 1 \leq j \leq p.
		\end{cases}
	\end{align}
	The parameter $ t \leq \min(n-1, p) $ restricts the $ \ell_0 $-norm of the columns of $ \bbet_{1:G} $ and thus the number of nonzero coefficients in each model. The parameter $ u \leq G $ restricts the $ \ell_0 $-norm of the rows of $ \bbet_{1:G} $ and thus the number of models that share any given predictor. Note that if $ u=G $, then \eqref{eq:BSpS} is equivalent to BSS in \eqref{BSS} for the same value of $ t $ and there is no diversity among the models. Hence, BSpS may be seen as a generalization of BSS to multiple models.
	The tuning parameters may be chosen in a data-driven manner, e.g. via CV. 
	
	While there are many proposals in the literature to obtain an optimal ensembling function \citep[see e.g.][]{breiman1996stacked}, the ensemble fit corresponding to the models $\hbbet^1,\dots,\hbbet^G$ selected by~\eqref{eq:BSpS} is  obtained by 
	\begin{align} \label{eq:ensemble_fit}
		\bbbet = \frac{1}{G} \sum_{g=1}^G \hbbet^g.
	\end{align}
	With this choice, the ensemble model remains an interpretable, sparse linear model similarly as for single model regularization methods, but in contrast to algorithmic ensemble methods. The ensemble model combines the information of the $G$ individual models, which each provide an explanation for the relationship between a subset of the predictors and the response.
	
	\subsection{Split Combinatorics} \label{sec:split_combinatorics}
	To solve the nonconvex problem in ~\eqref{eq:BSpS} exactly, all possible ways to construct $G$ subsets containing at most $t$ variables such that no variable appears more than $u$ times need to be considered. 
	The total number of possible splits of $ p $ variables into $ G $ groups, for $ p \geq G $, was derived by \cite{christidis2020split}. We extend their combinatorics result to the BSpS optimization problem \eqref{eq:BSpS} for the case without overlap between the models ($ u=1 $). Note that the computational problem for BSpS is even larger if predictors are allowed to be shared between groups ($ u>1 $).
	
	Let $ p_g $ be the number of variables in group $ g $, $ 1 \leq g \leq G $, and let $q = \sum_{g=1}^G p_g$.  
	Also let $h_i(p_{1},\dots,p_{G})$ be the number of elements in the sequence $p_{1},\dots,p_{G}$ that are equal to $ i $, $ 1 \leq i \leq t$. The number of possible splits of $p$ features into $G$ groups containing at most $ t $ variables is given by
	\begin{align} \label{eq:total_splits}
		\mathcal{T}(p, G, t)  =\sum_{p_{1}\leq \cdots\leq p_{G}\leq t} \binom{p}{q}\left[\frac{q!}{p_{1}! \dots p_{G}!} \prod_{i=1}^{t}\frac{1}{h_i(p_{1},\dots,p_{G})!}\right].
	\end{align}
	For example, $\mathcal{T}(15,3,10) =$ 171,761,951. 
	Thus, even for a relatively small number of predictor variables, the issue of computational infeasibility of BSpS becomes apparent and will be magnified further if $ t $ and $ u $ in \eqref{eq:BSpS} are chosen by CV.  
	
	\subsection{Related Work} \label{sec:multi_convex}
	
	\cite{christidis2020split} recently introduced the Split Regularized Regression (SplitReg) method which can be seen as a multi-convex relaxation of BSpS. While hard thresholds are used for BSpS in~\eqref{eq:BSpS}, soft thresholds are used in SplitReg which can be incorporated in the objective function more easily. 
	
	In detail, SplitReg is a minimizer $ \tilde{\bbet}_{1:G} = (\tilde{\bbet}^1, \dots, \tilde{\bbet}^G) \in \mathbb{R}^{p \times G} $ of an objective function of the form
	\begin{align} \label{eq:splitreg}
		\mathcal{J}\left(\by, \bX, \bbet^1, \dots, \bbet^g \right) = \sum_{g=1}^G \left\{\frac{1}{2n} \left\lVert \by - \bX \bbet^g \right\rVert_2^2 + \lambda_s P_s\left( \bbet^g \right) + \frac{\lambda_d}{2} \sum_{g\neq h}^G P_d\left(\bbet^h, \bbet^g\right) \right\},
	\end{align}
	where $ P_s $ and $ P_d $ are sparsity and diversity penalty functions. The tuning constants $\lambda_s, \lambda_d >0$ may be chosen e.g. by CV and control the magnitude of the sparsity and diversity penalties. 
	\cite{christidis2020split} proposed to use as sparsity and diversity penalties 
	\begin{align}
		P_s(\bbet) = \lVert \bbet \rVert_1 \quad
		\text{ and } \quad
		P_d(\bbet^g, \bbet^h) = \sum_{j=1}^p \left\lvert \bbet_j^g \right\rvert \left\lvert \bbet_j^h \right\rvert.
		\label{eq:splitreg_penalties}
	\end{align}
	Hence, $P_s(\bbet)$ is the Lasso penalty and similarly the diversity penalty $P_d(\bbet^g, \bbet^h)$ is an $\ell_1$-norm relaxation of the hard threshold in~\eqref{eq:BSpS}. 	The SplitReg objective function is multi-convex and can be solved efficiently via a block coordinate descent algorithm.
	\citet{christidis2020split} showed that with the penalties in~\eqref{eq:splitreg_penalties} the ensemble estimator given by  \eqref{eq:ensemble_fit} yields consistent predictions and has a fast rate of convergence. 

	Unlike the parameters $ t $ and $ u $ in BSpS, the parameters $ \lambda_s $ and $ \lambda_d $ do not directly control the number of predictors in each model and the number of models that can share any given predictor. In fact, there is theoretical and empirical evidence that such sparse relaxations can negatively affect variable selection performance \citep[see e.g.][]{van2009conditions, hazimeh2020fast}. Moreover,  in SplitReg the penalties \eqref{eq:splitreg_penalties} induce shrinkage of the coefficient estimates which may have a negative effect on prediction performance in high signal-to-noise scenarios \citep[see e.g.][]{hastie2020best}. 
	Therefore, we develop an algorithm to directly optimize BSpS in \eqref{eq:BSpS}. In the next Section we first generalize forward stepwise regression to the BSpS problem in \eqref{eq:BSpS} for the special case $ u=1 $, which then provides a starting point for our main algorithm in Section \ref{sec:PSGD}.
	
	\section{Initial Estimator} \label{sec:stepwise}

	To develop a fast algorithm that yields solutions for BSpS in \eqref{eq:BSpS} for the particular case of $ u=1 $ (i.e. when models are fully disjoint), we generalize forward stepwise regression to multi-model regression ensembles.
	
	
	For notational convenience, for any subset $ S \subseteq \{1, \dots, p\}$ we denote the cardinality of the set $ S $ by $ |S| $, $ \bbet_S \in \mathbb{R}^{|S|}$ denotes the subvector of $ \bbet \in \mathbb{R}^p$ with  element indices in $ S $, and $ X_S \in \mathbb{R}^{n \times |S|}$ denotes the submatrix of $ X $ with column indices in $ S $. Moreover, let $ I_n \in \mathbb{R}^{n \times n}$ denote the identity matrix of order $ n $ and $ F_{(d_1, d_2)}^{-1}(t) $ the quantile function of the $ F $-distribution with $ d_1 $ and $ d_2 $ degrees of freedom, respectively.

	At the start of the algorithm, all models are empty, i.e. they do not contain any predictor variables. The algorithm then iterates the following procedure until all models are saturated, which means that either the model contains $ n-1 $ predictors or there is no remaining candidate predictor providing a large enough improvement to the goodness-of-fit  of the model (i.e. with $p$-value below the threshold). 
	
	At each iteration, the candidate predictor that provides the largest improvement in goodness-of-fit of each unsaturated model  is identified by calculating for each candidate predictor its partial correlation $\tau_j^{(g)}$ with the response.
	The size of the improvement for this model is then measured by the $p$-value of the standard $F$-test for nested model comparison. 
	If the smallest $p$-value (i.e. the largest possible improvement) falls below a chosen threshold $ \gamma \in [0,1] $, then the corresponding model is updated by adding the identified predictor to its set of model predictors. This predictor is then also removed from the set of candidate predictors and thus can no longer be used in another model. This process is repeated until all $ G $ models are saturated.
	In the second step the Lasso is then applied to each model with tuning parameter $\lambda^{(g)}$ chosen by CV.  
	
	The proposed stepwise split regression algorithm, which we call Step-SplitReg,  is described in detail in Algorithm \ref{alg:stepwise_algo}. 
	%
	%
	An implemention of Algorithm \ref{alg:stepwise_algo} is available in \texttt{R} package \texttt{stepSplitReg} \citep{stepSplitReg} on CRAN \citep{CRAN}. 
	A  reference manual with the complete details of the package is available at \url{https://CRAN.R-project.org/package=stepSplitReg}.

	\begin{algorithm}[ht!]
		\caption{Stepwise Split (Regularized) Regression \label{alg:stepwise_algo}}
		\begin{algorithmic}[1]
			\Require{Design matrix $\bX \in \mathbb{R}^{n \times p}$, response vector $\by \in \mathbb{R}^n$,  number of models $ G \geq 2 $, and significance threshold $ \gamma \in (0,1) $.}
			\vspace{0.2cm}
			\Ensure{Set the set of candidate predictors $ J = \{1, \dots, p\} $ and $\gamma^*=0$. For each model ($ 1 \leq g \leq G $) set the set of predictors $ J^{(g)} = \emptyset $, the model saturation indicator $ T^{(g)} = \textsc{false} $, $ \hbbet^g = \mathbf{0} \in \mathbb{R}^p $  and the hat matrix $ H^{(g)} = \mathbf{0}  \in \mathbb{R}^{n\times n}$. }
			\Statex
			\State Repeat the following steps until $ \gamma^* \geq \gamma $ or $ T^{(g)}= \textsc{true} $ for all $ 1 \leq g \leq G $: \label{alg1:step1}
			\begin{enumerate}[label*=\footnotesize 1.\arabic*:]
				\item For each model $ g $ satisfying $ T^{(g)}=\textsc{false} $: 
				\begin{enumerate}[label*=\footnotesize \arabic*:]
					\item Identify candidate predictor maximizing decrease in residual sum of squares,
					\begin{align*}
						j^{(g)} = \argmax_{j \in J} \tau_j^{(g)}, \quad \tau_j^{(g)} = \frac{\by^T\left(I_n-H^{(g)}\right) \bx_j}{\bx_j^T\left(I_n-H^{(g)}\right) \bx_j}.
					\end{align*}
					\item Calculate the $ p $-value $ \gamma^{(g)} $ of predictor $ j^{(g)} $ in the enlarged model,
					\begin{align*}
						\gamma^{(g)} = 1 - F_{(1,|J^{(g)}|+1)}^{-1}\left(\frac{\left(|J^{(g)}|+1\right)\tau_{j^{(g)}}^{(g)}}{\left(\by - H^{(g)}X_{J^{(g)}}\right)^T\left(\by - H^{(g)}X_{J^{(g)}}\right) - \tau_{j^{(g)}}^{(g)}}\right).
					\end{align*}
					\item If $ \gamma^{(g)} \geq \gamma$ set $T^{(g)}=\textsc{true} $.
				\end{enumerate}
				\item Identify the unsaturated model $ g^* $ with the smallest $ p $-value $ \gamma^{(g^*)} $.  Set   $\gamma^*= \gamma^{(g^*)} $.
				\item If $ \gamma^* < \gamma $:
				\begin{enumerate}[label*=\footnotesize \arabic*:]
					\item Update the set of candidate predictors $ J = J \setminus \{j^{(g^*)}\} $ and the set of predictors for model $g^*$: $ J^{(g^*)} = J^{(g^*)} \cup \{j^{(g^*)}\} $.
					\item If $ |J^{(g^*)}| = n-1 $, set $T^{(g^*)}=\textsc{true} $. Otherwise, update the model hat matrix
					\begin{align*}
						H^{(g^*)} = X_{J^{(g^*)}} \left(X_{J^{(g^*)}}^T X_{J^{(g^*)}}\right)^{-1} X_{J^{(g^*)}}^T.
					\end{align*}
				\end{enumerate}         
			\end{enumerate}
			\State For each model $g=1,\dots,G$, set \label{alg1:step2}
			\begin{align*}
				\hbbet_{|J^{(g)}|}^g = \argmin_{\bbet \in \mathbb{R}^{|J^{(g^*)}|}} \lVert \by -\bX_{J^{(g)}} \bbet \rVert_2^2 + \lambda^{(g)} \lVert \bbet \rVert_1,
			\end{align*}
			where $\lambda^{(g)}$ is chosen by CV. 
			\Statex
			\State Return the sets of model predictors $ J^{(g)} $ and their coefficients $ \hbbet^{g} $, $ 1 \leq g \leq G $.
		\end{algorithmic}
	\end{algorithm}

	\section{Projected Subsets Gradient Descent Algorithm} \label{sec:PSGD}
	
	By adapting ideas from $\ell_0$-penalized optimization  for the BSS problem in \eqref{BSS} we develop an  algorithm to calculate approximate solutions for the BSpS problem~\eqref{eq:BSpS}. 
	
	\subsection{Solutions for Fixed $ t $ and $ u $}
	
	The least squares loss function used in \eqref{eq:BSpS} can be written as
	\begin{align} \label{eq:loss_function}
		\mathcal{L}_n\left(\bbet | \by, \bX \right) = \|\by - \bX \bbet\|^2,
	\end{align}
	with gradient given by
	\begin{align} \label{eq:gradient}
		\nabla_{\bbet} \mathcal{L}_n\left(\bbet | \by, \bX \right) = 2 \bX^T (\bX \bbet - \by).
	\end{align}
	Since 
	\begin{align*}
		&\left\Vert \nabla	\mathcal{L}_n\left(\bbet | \by, \bX \right) - \nabla \mathcal{L}_n\left(\tbbet | \by, \bX \right) \right\Vert_2\\
		= &\left\Vert 2\bX^T\left(\bX \bbet - \by \right) - 2 \bX^T\left(\bX \tbbet - \by \right) \right\Vert_2 \\
		= &\left\Vert 2 \bX^T \bX \left(\bbet-\tbbet \right) \right\Vert_2 \\
		\leq &\left\Vert 2 \bX^T\bX \right\Vert_2 \left\Vert \bbet - \tbbet \right\Vert_2 \\
		= &2 \left\Vert  \bX^T\bX \right\Vert_2 \left\Vert \bbet - \tbbet \right\Vert_2, 
	\end{align*}
	it follows that \eqref{eq:gradient} is Lipschitz continuous with Lipschitz constant $ \ell_{\bbet} = 2 \lVert \bX^T \bX \rVert_2$, where $ \lVert \bX^T \bX \rVert_2 $ is the spectral norm of $ \bX^T \bX $.
	Thus the loss function \eqref{eq:loss_function} is bounded from above by its quadratic approximation with Lipschitz constant $ \ell_{\bbet} $ \citep{boyd2004convex},
	\begin{align*}
		\mathcal{L}_n\left(\bbet | \by, \bX \right) &\leq \mathcal{L}_n\left(\tbbet | \by, \bX \right) + \nabla_{\bbet} \mathcal{L}_n\left(\tbbet | \by, \bX \right)^T \left(\bbet - \tbbet\right)+ \frac{1}{2} \ell_{\bbet} \big\lVert \bbet - \tbbet \big\rVert_2^2 \\
		&= \mathcal{L}_n^Q\left(\bbet | \by, \bX, \tbbet \right).
	\end{align*}
	
	For $1 \leq g \leq G$, define  $ S^{(g)} \subseteq J = \{1,\dots, p\} $ the subset of predictors that are used in at most $ u-1 $ models excluding model $ g $, i.e.
	\begin{align}
		S^{(g)} = \left\{j \in J: \sum_{\substack{h=1 \\ h \neq g}}^G \mathbb{I}\left(j \in J^{(h)}\right) \leq u-1\right\},
	\end{align}
	where $ J^{(g)}=\{j \in J: \hbbet_j^g \neq 0 \} $, $ 1 \leq g \leq G $, as defined in Section \ref{sec:stepwise}.
	Central to our algorithm is the projected subset operator, which we define for any vector $ v \in \mathbb{R}^p$ and some subset $ S  \subseteq J$ as
	\begin{align} \label{eq:projected_subset}
		\mathcal{P}(v; \, S, t) \in \argmin_{w \in \mathbb{R}^p} \; \lVert w - v \rVert_2^2 
		\quad \text{subject to} \quad 
		\begin{cases}
			\lVert w \rVert_0 \leq t, \text{ and} \\
			\{j \in J: w_j \neq 0 \} \subseteq S.
		\end{cases}
	\end{align}
	The operator $ \mathcal{P}(v; \, S, t) $ retains the $ t $ largest elements in absolute value of the vector $ v $ that belong to the set $ S $. It is a set-valued map since more than one possible permutation of the indices $ \{j \in J: j \in S\} $ may exist. 
	
	Starting from initial solutions $ \tbbet^1, \dots, \tbbet^G $, our algorithm applies projected subset gradient descent to each model cyclically until convergence. 
	Let $ \ell_{\bbet}^{(g)} = 2 \lVert \bX_{S^{(g)}}^T \bX_{S^{(g)}} \rVert_2 $, where $ \bX_{S^{(g)}} \in \mathbb{R}^{n \times |S^{(g)}|} $ is the submatrix of $ \bX $ with column indices in $ S^{(g)} $.
	Then for any $ L_{\bbet}^{(g)} \geq \ell_{\bbet}^{(g)} $ the updates in each iteration for model $ g $ 
	using subset $ S^{(g)} $ and model size $ t $ can be written as
	
	\begin{align*}
		\hbbet^g &\in \argmin_{\bbet^g \in \mathbb{R}^p} \; \mathcal{L}_n^Q\left(\bbet^g | \by, \bX, \tbbet^g \right) 
		\quad \text{subject to} \quad
		\begin{cases}
			\lVert \bbet^g \rVert_0 \leq t,  \text{ and} \\
			\{j \in J: \bbet_j^g \neq 0 \} \subseteq S^{(g)}.
		\end{cases} \\
		&= \argmin_{\bbet^g \in \mathbb{R}^p} \; \Bigg\lVert \bbet^g - \left(\tbbet^g - \frac{1}{L_{\bbet}^{(g)}}\nabla_{\bbet} \mathcal{L}_n\left(\bbet^g | \by, \bX \right)\bigg|_{\bbet=\tbbet^g}\right) \Bigg\rVert_2^2 \quad \text{subject to} \quad
		\begin{cases}
			\lVert \bbet^g \rVert_0 \leq t,  \text{ and} \\
			\{j \in J: \bbet_j^g \neq 0 \} \subseteq S^{(g)}.
		\end{cases} \\
		&=  \mathcal{P}\left(\tbbet^g - \frac{1}{L_{\bbet}^{(g)} } {\nabla}_{\bbet} \mathcal{L}_n\left(\bbet^g|\by, \bX\right)\bigg|_{\bbet^g=\tbbet^g}; \, S^{(g)}, t \right).
	\end{align*}

	Note that for each model the iterative algorithm produces a sequence of converging solutions  \citep[Proposition~6]{bertsimas2016best}. In particular, the iterative algorithm converges to an $ \epsilon $-optimal stationary point in $ O(1/\epsilon) $ iterations \citep[Theorem~3.1]{bertsimas2016best}, i.e. for any $ \epsilon >0 $ we have that
	\begin{align*}
		\left\lVert \hbbet^g - \mathcal{P}\left(\hbbet^g - \frac{1}{L_{\bbet}^{(g)} } {\nabla}_{\bbet} \mathcal{L}_n\left(\bbet^g|\by, \bX\right)\bigg|_{\bbet^g=\hbbet^g}; \, S^{(g)}, t \right) \right\rVert_2^2 \leq \epsilon.
	\end{align*}
	After convergence is reached, its set of predictor variables $ J^{(g)} = \{j \in J: \hbbet_j^g \neq 0 \} $ is updated and the final model coefficients are computed. 
	
	The projected subsets gradient descent (PSGD) algorithm is described in detail in Algorithm \ref{alg:projected_algo}. 
	In the optional step \ref{alg2:step3} of the algorithm, local combinatorial searches using random permutations of the group order are performed to improve on the projected subset gradient descent solution from step \ref{alg2:step2}. A total of $ G! $ new starting points are available for the local combinatorial searches. Based on our numerical experiments, a poor initial solution for step \ref{alg2:step2} increases the need for step \ref{alg2:step3}  to obtain better solutions which come with a high computational cost. However, a carefully designed algorithm that generates good initial solutions for Algorithm \ref{alg:projected_algo} alleviates the need for the random permutations of the models (step \ref{alg2:step3}) when cyclically updating the models in step \ref{alg2:step2}.
	
	\begin{algorithm}[ht!]
		\caption{\label{alg:projected_algo}Projected Subsets Gradient Descent (PSGD)}
		\begin{algorithmic}[1]
			\Require{Design matrix $\bX \in \mathbb{R}^{n \times p}$, response vector $\by \in \mathbb{R}^n$,  initial solutions  $ \tbbet^1, \dots, \tbbet^G $, sparsity and diversity tuning parameters $ t $ and $ u $, tolerance parameter $\epsilon>0$, and (optional) number of local combinatorial searches $ C $.}
			\Statex
			\State Initialize the sets of model predictors $ J^{(g)} = \{j \in J: \tbbet_j^g \neq 0 \} $, $ 1 \leq g \leq G $. \label{alg2:step1}
			\Statex
			\State Repeat the following steps for each model $ g $, $ 1 \leq g \leq G $: \label{alg2:step2}
			\begin{enumerate}[label*=\footnotesize 1.\arabic*:]
				\item[\footnotesize 2.1:] Update the allowed predictors
				\begin{align*}
					S^{(g)} = \left\{j \in J: \sum_{h \neq g}^G \mathbb{I}\left(j \in J^{(h)}\right) \leq u-1\right\},
				\end{align*}
				and the Lipschitz constant $ \ell_{\bbet}^{(g)} = 2 \lVert \bX_{S^{(g)}}^T \bX_{S^{(g)}} \rVert_2  $.
				\item[\footnotesize 2.2:] Update $ \tbbet^g $ as
				\begin{align*}
					\hbbet^g \in \mathcal{P}\left(\tbbet^g - \frac{1}{L_{\bbet}^{(g)} } {\nabla}_{\bbet} \mathcal{L}_n\left(\bbet^g|\by, \bX\right)\Big\vert_{\bbet^g = \tbbet^g}; \, S^{(g)}, t \right),
				\end{align*}
				with $L_{\bbet}^{(g)} \geq \ell_{\bbet}^{(g)} $   until $\mathcal{L}_n(\tbbet^g|\by, \bX) - \mathcal{L}_n(\hbbet^g|\by, \bX) \leq \epsilon$.
				\item[\footnotesize 2.3:] Update the model predictors $ J^{(g)} = \{j \in J: \hbbet_j^g \neq 0 \} $.
				\item[\footnotesize 2.4:] Compute the final model coefficients
				\begin{align*}
					\hbbet^g = \argmin_{\bbet^g \in \mathbb{R}^p} \; \mathcal{L}_n\left(\bbet^g | \by, \bX\right) \quad \text{subject to} \quad \beta_j^g = 0, j \notin J^{(g)}.
				\end{align*}
			\end{enumerate}
			\State (Optional) Repeat the following steps $ C $ times: \label{alg2:step3}
			\begin{enumerate}[label*=\footnotesize 2.\arabic*:]
				\item[\footnotesize 3.1:] Draw a random permutation $  (\omega(1), \dots, \omega(G)) $ of $ (1, \dots, G) $. 
				\item[\footnotesize 3.2:] Repeat step \ref{alg2:step2} using the new order $  (\omega(1), \dots, \omega(G)) $  and the initial solutions.
				\item[\footnotesize 3.3:] If the new solution improves on the incumbent solution update the sets of model  predictors $ J^{(g)} $ and their coefficients $ \hbbet^g $, $ 1 \leq g \leq G $, with the new solutions. Otherwise keep the old solutions.
			\end{enumerate}
			\Statex
			\State Return the sets of model predictors $ J^{(g)} $ and their coefficients $ \hbbet^g $, $ 1 \leq g \leq G $. 
		\end{algorithmic}
	\end{algorithm}
	
	\subsection{Selection of Tuning Parameters}
	
	To generate good initial solutions for Algorithm \ref{alg:projected_algo}, in Algorithm \ref{alg:incrementing_projected_algo} we progressively reduce the diversity of the models  until $ u =G $ in \eqref{eq:BSpS}. Algorithm \ref{alg:incrementing_projected_algo} starts with $u=1$ for which  the initial solution, i.e. initial sets of model predictors and corresponding coefficient estimates, is obtained by Algorithm \ref{alg:stepwise_algo}. Algorithm \ref{alg:projected_algo} is then applied with this initial solution to obtain the final solution for $u=1$. For $u=2,\dots,G$, Algorithm \ref{alg:projected_algo} is then successively applied with the solution for $u-1$ as initial solution for the next value of $u$.
	Even without the optional local combinatorial searches in step \ref{alg2:step3} of Algorithm \ref{alg:projected_algo}, Algorithm \ref{alg:incrementing_projected_algo} produces competitive solutions in terms of minimizing the objective function of BSpS in \eqref{eq:BSpS}  based on our numerical experiments. Hence, for the analysis of simulated and real data in the remainder of this article we use Algorithm \ref{alg:incrementing_projected_algo} without the optional last step of Algorithm \ref{alg:projected_algo}.
	
	The values of the sparsity and diversity tuning parameters in \eqref{eq:BSpS}, $ t $ and $ u $ respectively, need to be determined from the training data. We use 5-fold CV on grids of candidate values for $ t $ and $ u $ and select the values which minimize the CV MSPE. For a fixed sparsity level $ t $, Algorithm \ref{alg:incrementing_projected_algo} is well suited to generate solutions for a grid of candidates for $ u $ by using warm starts. This process is then repeated for every candidate $ t $, which can be any subset of  $ \{1, \dots, n-1\} $. Note that Algorithm  \ref{alg:stepwise_algo} does not depend on the value of $t$, so Step 1 of Algorithm \ref{alg:incrementing_projected_algo} only needs to be executed for the first value of $t$.

	\begin{algorithm}[ht!]
		\caption{Decrementing Diversity PSGD \label{alg:incrementing_projected_algo}}
		\begin{algorithmic}[1]
			\Require{Design matrix $\bX \in \mathbb{R}^{n \times p}$, response vector $\by \in \mathbb{R}^n$, Lipschitz constant $ L_{\bbet} $, maximum model size $ t $, and tolerance parameter $\epsilon>0$.}
			\Statex
			\State Set $ u=1 $ and use Algorithm \ref{alg:stepwise_algo} to initialize the sets of model predictors $ J^{(g)}(1) $  and their coefficients $ \hbbet^g(1) $, $ 1 \leq g \leq G $. 
			\Statex
			\State Using Algorithm \ref{alg:projected_algo} update $ J^{(g)}(1) $ and $ \hbbet^g(1) $ with the solution from step 1 as initial solution.
			\Statex
			\State For $ u=2,\dots, G $ repeat the following step:
			\begin{enumerate}[label*=\footnotesize 2.\arabic*:]
				\item Compute  $ J^{(g)}(u) $ and  $ \hbbet^g(u) $ using Algorithm \ref{alg:projected_algo} with initial solutions $ J^{(g)}(u-1) $ and $ \hbbet^g(u-1) $, $ 1 \leq g \leq G $.
			\end{enumerate}
			\Statex
			\State Return the sets of model predictors $ J^{(g)}(u) $ and their coefficients $ \hbbet^g(u) $, $ 1 \leq u,g \leq G $. 
		\end{algorithmic}
	\end{algorithm}

	
	An implementation of the PSGD Algorithms \ref{alg:projected_algo},  \ref{alg:incrementing_projected_algo} and the CV procedure is available in \texttt{R} package \texttt{PSGD} \citep{PSGD} on CRAN. Multithreading with OpenMP \citep{chandra2001parallel} is available in the package to reduce the computational cost of the method. 
	A reference manuel with the complete details of the package is available at \url{https://CRAN.R-project.org/package=PSGD}.

	\section{Simulation Study} \label{sec:simulation}
	
	For each Monte Carlo replication, we generate data from the linear
	model
	\begin{equation*}
		y_{i} = \mathbf{x}_{i}^{\prime} \boldsymbol{\beta}_{0} + \sigma \epsilon_{i},
		\quad 1\leq i \leq n,
	\end{equation*}
	where the $\mathbf{x}_{i}\in\mathbb{R}^{p}$ are multivariate normal with
	zero mean and correlation matrix $\boldsymbol{\Sigma} \in \mathbb{R}^{p \times p}$ and the $\epsilon_{i}$
	are standard normal. We set $n=50$ and  $p$ is either $150$ or $500$.
	For each $p$, we consider the proportion of active (i.e. nonzero) variables 
	$\zeta \in \{0.1, 0.2, 0.4\}$ . 
	
	The $p_0=[p \zeta]$ nonzero elements of the $p$-dimensional vectors $\boldsymbol{\beta}_0$ are randomly generated as described in \cite{SIS}, i.e. nonzero coefficients are generated according to $ (-1)^u (a + |z|) $, with $a = 5 \log n/\sqrt{n}$ and where $u$ is drawn from a Bernoulli distribution with parameter 0.2 and $ z $ is drawn from the standard Gaussian distribution. 
	We consider two different scenarios for $\boldsymbol{\Sigma}$.\\
	
	\noindent
	\textbf{Scenario 1:}
	\begin{equation*}
		\Sigma_{i,j}=  
		\begin{cases}
			1 & \text{if } i=j, \\ 
			\rho & \text{if } i\neq j. \\ 
		\end{cases}
	\end{equation*}
	
	\noindent
	\textbf{Scenario 2:} 
	\begin{equation*}
		\Sigma_{i,j}=  
		\begin{cases}
			1 & \text{if } i=j, \\ 
			\rho & \text{if } 1\leq i,j \leq  p_0, i\neq j, \\ 
			0 & \text{otherwise}.
		\end{cases}
	\end{equation*}
	
	In Scenario 1, all the predictors are correlated among each other. In Scenario 2, the active variables are only correlated with each other.
	For both scenarios we consider the values $\rho \in \{0.2, 0.5, 0.8\}$.
	Finally, $\sigma$ is chosen to give a desired signal to noise ratio (SNR), defined 
	as $\text{SNR} = {\bbet_{0}^{\prime} \boldsymbol{\Sigma} \bbet_{0}}/{\sigma^2}.$
	We consider SNRs of 1, 3 and 5.
	We report results for both scenarios across all considered sparsity levels, correlations, SNRs and dimensions $p$.
	
	\subsection{Methods}
	
	Our simulation study compares the prediction accuracy of eleven methods. In particular, we consider four sparse regression methods, their analogous split regression methods, and three ``blackbox" regression ensemble methods. 
	All computations were carried out in \texttt{R} using the default settings for the tuning parameters in the implementations of the methods.
	\begin{enumerate}
		\item[1.] \textbf{Stepwise} forward regression, computed using the \texttt{lars} package \citep{lars}.
		\item[2.] \textbf{Lasso}, computed using the \texttt{glmnet} package \citep{glmnet}. 
		\item[3.] Elastic Net (\textbf{EN}) with $\alpha=3/4$ for the $ \ell_1$-$\ell_2 $ mixing parameter, computed using the \texttt{glmnet} package \citep{glmnet}.
		\item[4.] \textbf{Fast-BSS}, computed using the \texttt{L0Learn}  package \citep{L0Learn} with the $ \ell_0$-$\ell_2 $ penalty option.  
		\item[5.] \textbf{Step-SplitReg}, computed using the \texttt{stepSplitReg} package with a custom Lasso fit for each model. 
		\item[6.] \textbf{SplitReg-Lasso}, computed using the \texttt{SplitReg} package \citep{SplitReg}.
		\item[7.] \textbf{SplitReg-EN}  with $\alpha=3/4$ for the $ \ell_1$-$\ell_2 $ mixing parameter, computed using the \texttt{SplitReg} package.
		\item[8.] \textbf{Fast-BSpS}, computed using the \texttt{PSGD} package. 
		\item[9.] {Random GLM} (\textbf{RGLM}) \citep{random_glm_paper}, computed using the \texttt{RGLM} package \citep{randomGLM}.
		\item[10.] {Random Forest} (\textbf{RF}), computed using the \texttt{randomForest} package \citep{randomForest}.
		\item[11.] {Extreme Gradient Boosting} (\textbf{XGB}) \citep{chen2016xgboost}, computed using the \texttt{xgboost} package \citep{xgboost}.		
	\end{enumerate}
	
	For a fast computation of BSS, we use the state-of-the-art method of \cite{hazimeh2020fast}. \cite{L0Learn} recommend to combine $\ell_0$ regularization with shrinkage-inducing penalties to avoid overfitting and improve predictive performance, and thus we use the $\ell_0$-$\ell_2$ combination of penalties. For the four split regression methods, we use $ G = 5 $ models. For RGLM and RF we consider ensembles based on $ G = 5 $ models as well as ensembles based on their  default number of models which is $ G=100 $ and $ G=500 $, respectively. The number of models of XGB are chosen by its default CV procedure in \texttt{xgboost}.
	To reduce the computational burden of the PSGD algorithm in our large simulation study, we use the grids $ u \in \{1, 2, 3, 4, 5\} $ and $ t \in \{0.3n, 0.4n, 0.5n\} = \{15, 20, 25\} $ in the CV procedure of BSpS. 
	
	\subsection{Performance Measures}
	For each configuration, we randomly generate $ N = 50 $ training and test sets and for each
	of the methods measure average performance on the test sets. 
	In each replication of a particular configuration, the training set is used to fit the methods and a large independent test set of size $ m= $ 2,000 is used to compute the MSPE. The reported MSPEs  are relative to the irreducible error $\sigma^2$, hence the best possible result is $1$. We also report recall (RC) and precision (PR), which for a parametric method are defined as
	\begin{align*}
		\text{RC} = \frac{\sum_{j=1}^p\mathbb{I}(\beta_j\neq 0, \hat{\beta}_j\neq0)}{\sum_{j=1}^p\mathbb{I}(\beta_j\neq 0)}, \quad \text{PR} = \frac{\sum_{j=1}^p\mathbb{I}(\beta_j\neq 0, \hat{\beta}_j\neq0)}{\sum_{j=1}^p\mathbb{I}(\hat{\beta}_j\neq0)},
	\end{align*} 
	where $ \bbet $ and $ \boldsymbol{\hat{\beta}} $ are the true and estimated regression coefficients, respectively. 
	For the split regression methods and RGLM, the coefficients of the ensemble model given by \eqref{eq:ensemble_fit} are used to compute recall and precision.  For the tree-based ensemble methods  RF and XGB, the RC and PR are computed by identifying the predictors used in the trees of the ensemble. We do not report RC and PR of RGLM and RF when their default number of models are used because in this case their recall tends to be  always close to 1 while their precision equals approximately the proportion of active variables $\zeta$. 
	Note that large values of both RC and PR are desirable.

	\subsection{Results}
	To summarize the results, we ranked the 11 methods for each simulation setting and performance measure from best (rank 1) to worst (rank 11). 
	For each performance measure we report in Table \ref{tab:ranks_sim} the average rank of each method across all simulation settings. The best two ranks for each performance measure are in bold.
	The detailed results of the simulation study are available in the supplementary material.
	
	In terms of MSPE, Fast-BSpS had the best average rank for both $ p=150 $ and $ p=500 $, whereas SplitReg-EN had the second best performance in both cases. RGLM-100 had the best overall MSPE rank out of the black box methods. However, its performance deteriorated to the worst average rank when the number of models is set equal to $ G=5 $, the same as for Fast-BSpS. In Section \ref{sec:empirical_comparison}, we investigate this phenomenon in more detail by studying the effect of the number of models on Fast-BSpS and RGLM. Step-SplitReg was not competitive in terms of MSPE compared to  Fast-BSpS or the SplitReg methods. However it does outperform the  single-model stepwise methods consistently in terms of MSPE. 
	
	In terms of RC, Fast-BSpS had the second best rank overall, only beaten slightly by RGLM-5. However, RGLM-5 had the third worst overall rank in PR, whereas Fast-BSpS had the best PR rank for $ p=500 $ and the second best overall PR rank among all ensemble methods.
	
	\begin{table}[ht!]
		\centering
		\caption{Average rank of the methods over the scenarios, correlations, SNRs and sparsity levels for $(p,n)=(500,50)$ and $(p,n)=(150,50)$. The last column contains the overall rank over both combinations of $(p,n)$. \label{tab:ranks_sim}} 
		\extrarowsep =2pt
		\begin{tabu}{lrrrrrrrrr}
			\toprule & \multicolumn{3}{c}{$\mathbf{\boldsymbol{p}=500}$} & \multicolumn{3}{c}{$\mathbf{\boldsymbol{p}=150}$} & \multicolumn{3}{c}{\textbf{Overall Rank}} \\ \cmidrule(lr){1-1} \cmidrule(lr){2-4} \cmidrule(lr){5-7} \cmidrule(lr){8-10} \textbf{Method} & \textbf{MSPE} & \textbf{RC} & \textbf{PR} & \textbf{MSPE} & \textbf{RC} & \textbf{PR} & \textbf{MSPE} & \textbf{RC} & \textbf{PR} \\ \cmidrule(lr){1-1} \cmidrule(lr){2-4} \cmidrule(lr){5-7} \cmidrule(lr){8-10}
			\addlinespace[0.25cm]
			Stepwise & 12.06 & 11.00 & \textbf{3.87} & 11.17 & 11.00 & \textbf{3.09} & 11.62 & 11.00 & \textbf{3.48} \\ 
			Lasso & 7.20 & 9.81 & 4.17 & 6.50 & 9.78 & \textbf{3.67} & 6.85 & 9.80 & \textbf{3.92} \\ 
			EN & 6.17 & 8.81 & 4.13 & 5.93 & 8.72 & 4.30 & 6.05 & 8.77 & 4.21 \\ 
			Fast-BSS & 4.81 & 6.89 & 6.02 & 5.52 & 6.75 & 4.93 & 5.16 & 6.82 & 5.47 \\ 
			\addlinespace[0.25cm]
			Step-SplitReg & 9.07 & \textbf{1.85} & 10.26 & 6.96 & 5.21 & 8.96 & 8.02 & 3.53 & 9.61 \\ 
			SplitReg-Lasso & 3.57 & 5.06 & 6.09 & 3.33 & 5.55 & 5.00 & 3.45 & 5.30 & 5.54 \\ 
			SplitReg-EN & \textbf{2.85} & 3.89 & 5.57 & \textbf{2.74} & 4.60 & 5.41 & \textbf{2.80} & 4.25 & 5.49 \\ 
			Fast-BSpS & \textbf{2.56} & 3.56 & \textbf{3.20} & \textbf{2.09} & \textbf{2.28} & 5.78 & \textbf{2.33} & \textbf{2.92} & 4.49 \\ 
			\addlinespace[0.25cm]
			RGLM-5 & 12.24 & \textbf{3.24} & 8.46 & 12.69 & \textbf{1.46} & 9.50 & 12.46 & \textbf{2.35} & 8.98 \\ 
			RGLM-100 & 3.63 & $-$ & $-$ & 6.50 & $-$ & $-$ & 5.06 & $-$ & $-$ \\ 
			RF-5 & 10.02 & 7.65 & 10.15 & 10.30 & 6.13 & 10.67 & 10.16 & 6.89 & 10.41 \\ 
			RF-500 & 5.69 & $-$ & $-$ & 5.83 & $-$ & $-$ & 5.76 & $-$ & $-$ \\ 
			XGB & 11.13 & 4.24 & 4.07 & 11.44 & 4.52 & 4.70 & 11.29 & 4.38 & 4.38 \\ 
			\addlinespace[0.25cm]
			\bottomrule
		\end{tabu}
	\end{table}
	
\begin{figure}[ht!]
		\centering
		\includegraphics[width=16cm]{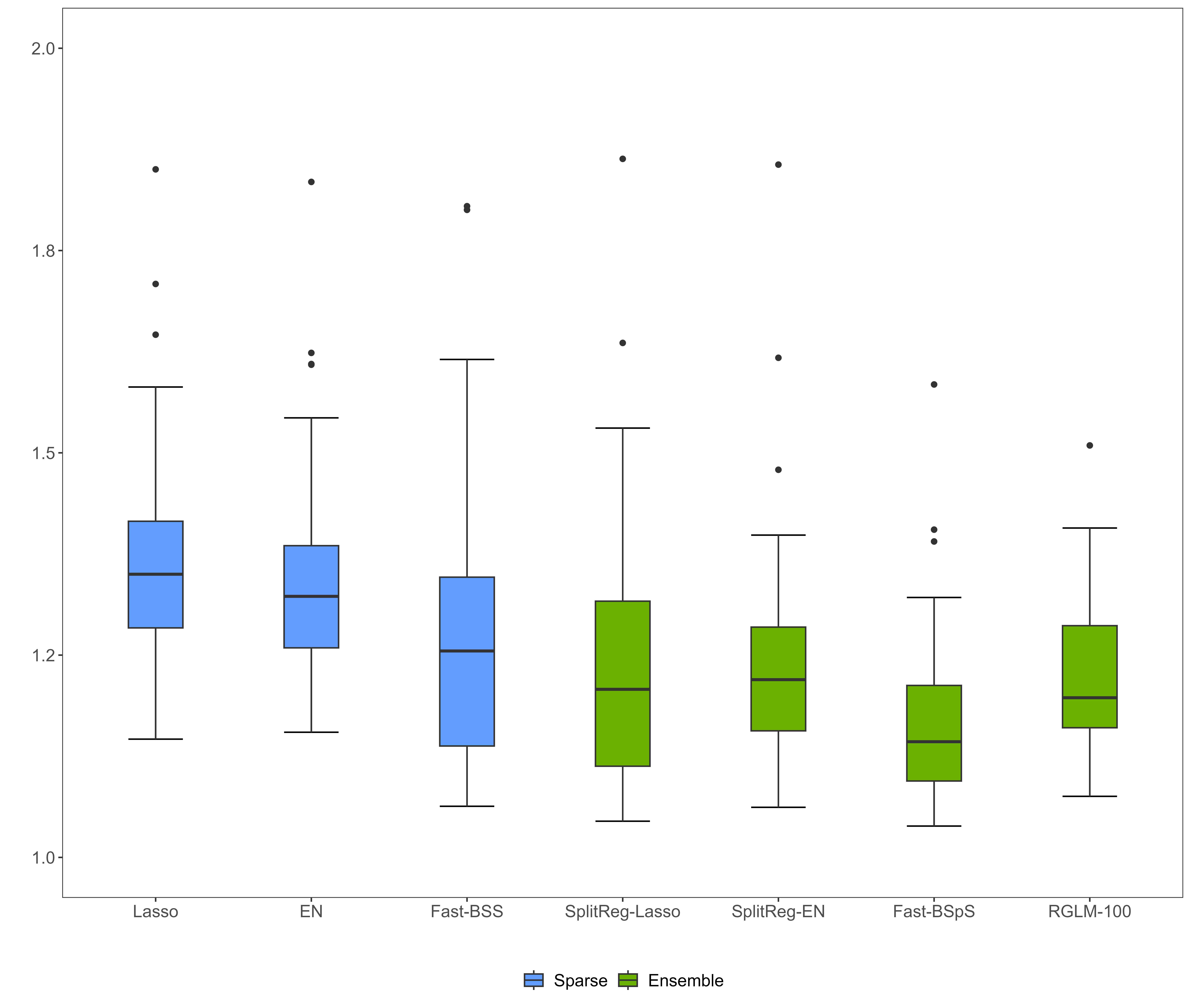}
		\caption{MSPEs of the sparse and ensemble regression methods over $ N=50 $ random training sets under Scenario 2 with $ \rho =0.5 $, $ p=\text{500} $, $ n=50 $, $ \text{SNR}=5 $ and $\zeta=0.4$.}
		\label{fig:BSpS_Simulation}
	\end{figure}

\begin{figure}[ht!]
		\centering
		\includegraphics[width=16cm]{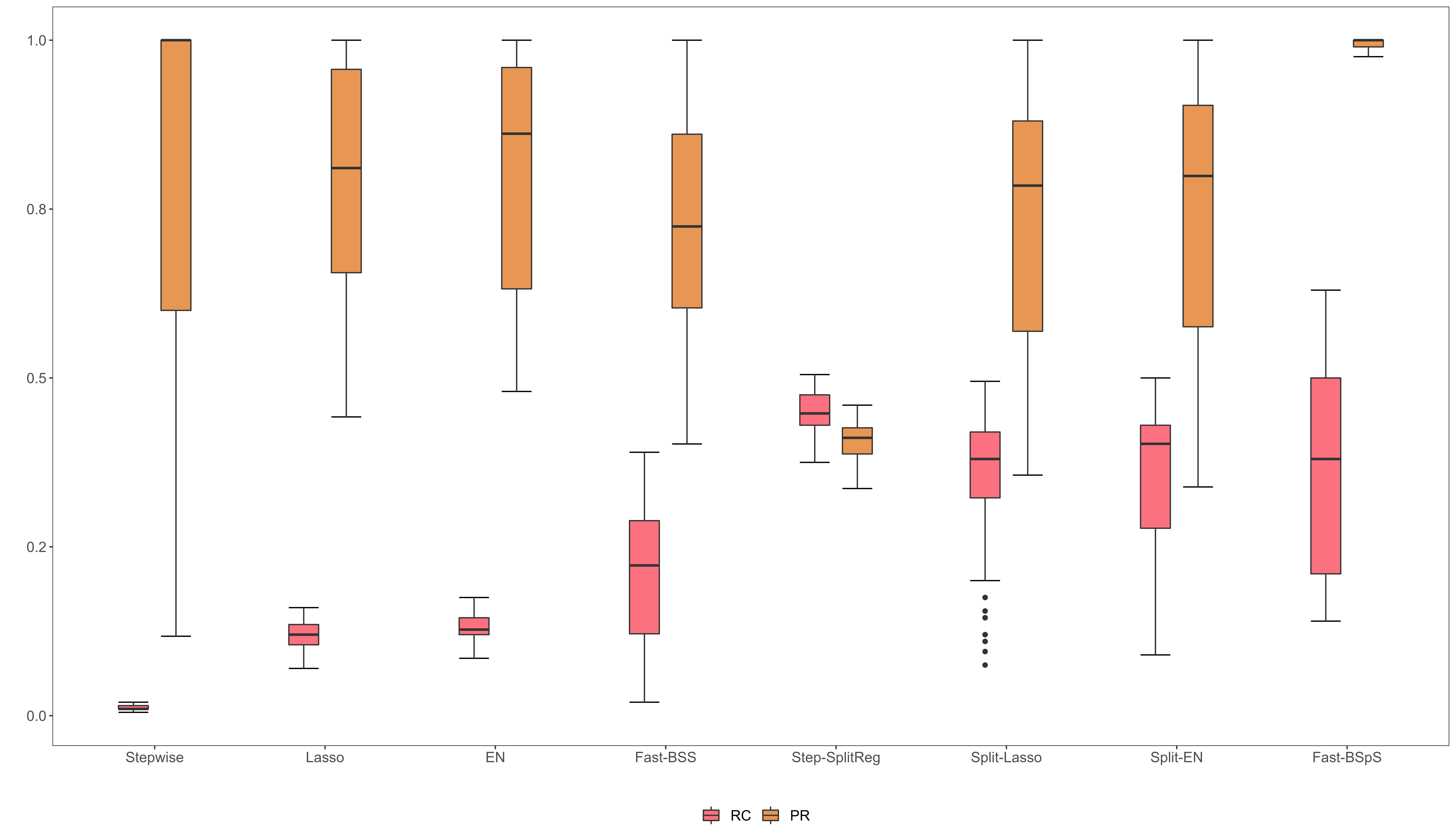}
		\caption{RC and PR of the sparse and split regression methods over $ N=50 $ random training sets under Scenario 2 with $ \rho =0.5 $, $ p=\text{500} $, $ n=50 $, $ \text{SNR}=5 $ and $\zeta=0.4$.}
		\label{fig:BSpS_Simulation_RCPR}
	\end{figure}

	In Figure \ref{fig:BSpS_Simulation} we plot the MSPEs of the Lasso, EN, Fast-BSS, SPlitReg-Lasso, SplitReg-EN, Fast-BSpS, and RGLM-100 under Scenario 2 with $ p=500 $,  $ \rho=0.5 $, $ \text{SNR}=5 $ and $\zeta=0.4$. We omit the other methods from this plot as they were not competitive in this scenario. Note that the third quartile of Fast-BSpS corresponds approximately to the median of its two closest competitors, Split-EN and RGLM-100. In Figure \ref{fig:BSpS_Simulation_RCPR} we plot the corresponding RC and PR of the sparse and split regression methods. We exclude the blackbox ensemble methods since their recall tends to be close to 1 and their precision close to the proportion of active variables, as mentioned previously. The median RC of Fast-BSpS is  similar to that of the multi-convex relaxations Split-Lasso and Split-EN. However, the PR of Fast-BSpS is clearly superior to the PR of all other methods, including Split-Lasso and Split-EN.
	
	\section{Insights in Fast-BSpS Performance} \label{sec:empirical_comparison}
	
	As discussed in Section~\ref{sec:ensembles} the Bias-Variance-Covariance (BVC) tradeoff in~\eqref{eq:MSPE_ensemble}-\eqref{eq:MSPE_variance} is a fundamental concept in understanding the behavior of ensemble methods. 
	The average bias and variance of the models in an ensemble and their covariances are crucial concepts when analyzing the performance of machine learning algorithms and were extensively discussed in the seminal paper by \cite{geman1992neural}, as well as in subsequent works by \cite{ueda1996generalization}, \cite{breiman1996bagging}, and \cite{hastie2009elements}.
	
	To gain insight into our method's performance, we empirically evaluated the bias, variance, and covariance components of our ensemble method. This analysis helps to understand how our method balances these components to reach an excellent prediction performance. We further explore the accuracy-diversity tradeoff by adjusting the number of models in our ensemble method for various simulation settings. By comparing accuracy and diversity metrics, we aim to understand how model diversity influences overall ensemble performance and its relationship with the BVC tradeoff. Furthermore, our simulations investigate the variable selection behavior of Fast-BSpS and analyze how the number of models affects this behavior.
	
	\subsection{Bias-Variance-Covariance Tradeoff} \label{sec:BVC}
	
	To evaluate the  BVC tradeoff of Fast-BSpS and RGLM we simulated $ N=50 $ training sets using Scenario 2 in the simulation study of Section \ref{sec:simulation}. We set $ p=500 $, proportion of active variables $ \zeta \in \{0.1, 0.2, 0.4\}$,	 
	correlation parameter $ \rho=0.5 $, and SNR $=$ 3. We again use training sets of size $ n=50 $ to fit the models and test sets of size $m = \text{2,000}$ to approximate $\overline{\text{Bias}}$, $\overline{\text{Var}}$ and $\overline{\text{Cov}}$ in \eqref{eq:MSPE_ensemble}-\eqref{eq:MSPE_variance}.
	To study the  BVC tradeoff, we include in our comparison Fast-BSpS with $G=5$ and RGLM with  $G=100$  (the default for RGLM).
	We also include Fast-BSS in the simulation study as a benchmark to assess the bias and variance of Fast-BSpS. 
	
	The results are reported in Table \ref{tab:BVC_Groups}. Each column shows values relative to the lowest achieved metric; thus, the best performing method is indicated by an entry of 1.00. For instance, a value of 57.84 for RGLM-100 in the $\overline{\text{Bias}}$ column at $\zeta=0.1$ indicates that RGLM-100's average individual model bias was more than fifty-seven times higher than that of Fast-BSpS, which achieved the lowest average individual model bias. From these results, it is evident that Fast-BSpS consistently achieves the lowest average individual model bias and effectively controls individual model variance, typically maintaining it within a threefold range compared to the single-model sparse method Fast-BSS, which focuses solely on variance control. However, RGLM generally minimizes average pairwise covariance. Despite this, RGLM's individual models exhibit notably high bias (often exceeding Fast-BSpS by over 100 times) and significantly higher average individual model variance, more than  20 times the average individual model variance of Fast-BSpS. Interestingly, Fast-BSpS also shows competitive average pairwise covariance with RGLM, never exceeding twice the value achieved by RGLM.
	
	\begin{table}[ht!]
		\centering
		\caption{MSPE, ${\overline{\text{MSPE}}}$ and $\overline{\text{Cor}}$ of Fast-BSpS and RGLM as a function of the number of models under Scenario 2 with $ \rho=0.5 $ and $\text{SNR} =3 $. \label{tab:BVC_Groups}} 
		\extrarowsep =2pt
		\begin{tabular}{lrrrrrrrrr}
			\toprule
			& \multicolumn{3}{c}{$\mathbf{\boldsymbol{\zeta}=0.1}$} & \multicolumn{3}{c}{$\mathbf{\boldsymbol{\zeta}=0.2}$} & \multicolumn{3}{c}{$\mathbf{\boldsymbol{\zeta}=0.4}$}\\ \cmidrule(lr){1-1} \cmidrule(lr){2-4} \cmidrule(lr){5-7} \cmidrule(lr){8-10}  \textbf{Method} & $\mathbf{\overline{Bias}}$ & $\mathbf{\overline{Var}}$ & $\mathbf{\overline{Cov}}$ & $\mathbf{\overline{Bias}}$ & $\mathbf{\overline{Var}}$ & $\mathbf{\overline{Cov}}$ & $\mathbf{\overline{Bias}}$ & $\mathbf{\overline{Var}}$ & $\mathbf{\overline{Cov}}$  \\ \cmidrule(lr){1-1} \cmidrule(lr){2-4} \cmidrule(lr){5-7} \cmidrule(lr){8-10} 
			\addlinespace[0.25cm]
			Fast-BSS & 34.00 & 1.00 & $ - $ & 1.57 & 1.00 & $ - $ & 3.66 & 1.00 &$ - $  \\ 
			\addlinespace[0.25cm]
			Fast-BSpS & 1.00 & 3.03 & 1.81 & 1.00 & 3.09 & 1.48 & 1.00 & 3.08 & 1.28 \\ 
			\addlinespace[0.25cm]
			RGLM-100 & 57.84 & 23.83 & 1.08 & 1.65 & 21.81 & 1.00 & 2.47 & 25.15 & 1.00 \\ 
			\addlinespace[0.25cm]
			\bottomrule
		\end{tabular}
	\end{table}
	
	The results show that Fast-BSpS  has the ability to manage well the BVC tradeoff. The two tuning parameters $u$ and $t$ in the BSpS objective function~\eqref{eq:BSpS} are crucial for controlling this tradeoff.  
	First, our method is able to control model diversity through the parameter $u$. By limiting the number of times a predictor can be used across the models in the ensemble, we enforce a diversity constraint that prevents too much reliance on any single predictor. This reduces covariance and mitigates the risk of overfitting, leading to better generalization. Additionally, our method's flexibility in setting the parameter $t$ allows for precise control on the complexity of each model in the ensemble. By tuning $t$, we can manage the tradeoff between bias and variance of the individual models: smaller model sizes (lower $t$) reduce variance but may increase bias, while larger model sizes (higher $t$) decrease bias at the cost of increased variance. Empirical results (see Table~\ref{tab:MSPE_Groups}) show that as more models are used, covariance is further reduced, which is crucial because, according to the BVC tradeoff \eqref{eq:MSPE_ensemble}-\eqref{eq:MSPE_variance}, with more models, covariance becomes more important while individual model variance becomes less important. Overall, our method effectively navigates these tradeoffs to achieve excellent performance. 
	
	In RGLM, bagging and the random subspace method are used to create $ G $ bags comprised of different samples and predictors. In each bag a subset of the predictors is then retained based on a measure of correlation with the response, and forward selection is applied to this subset.
	To generate a collection of diverse models 	RGLM thus randomly assigns candidate predictors to individual models. This results in models that are individually weak and not built to achieve a good accuracy-diversity balance given the number of models $ G $ that is used in the algorithm. From the bias-variance-covariance decomposition in \eqref{eq:MSPE_ensemble}-\eqref{eq:MSPE_variance} it can be seen that if the individual models are weak, a large number of diverse models is needed so that the covariance term can dominate the variance term in the variance of the ensemble and thus the ensemble achieves a good prediction accuracy.
	
	\subsection{Accuracy-Diversity Tradeoff}
	In the previous section, we showed that by controlling bias, variance and covariance of the models in the ensemble  Fast-BSpS 
	manages to find a good balance between individual model accuracy and diversity, leading to excellent ensemble accuracy. 
	The need to balance the accuracy of individual models with their diversity to achieve good ensemble performance is well-known in the ensemble learning literature \citep{kuncheva2003measures, brown2005managing}. 
	We now further investigate this accuracy-diversity tradeoff when the number of models in the ensemble is varied.
	
	We use the same simulation setup as in the BVC study. 
	To study the effect of the number of models on the performance of Fast-BSpS and RGLM, we apply these methods with the number of models $ G \in \{2,3,4,5\} $, as well as $ G=100 $ (the default) for RGLM for the same settings as in section \ref{sec:BVC}.  In Table \ref{tab:MSPE_Groups}, we report the ensemble Mean Squared Prediction Error (MSPE), the average MSPE of the individual models (${\overline{\text{MSPE}}}$), and their average correlation ($\overline{\text{Cor}}$). The MSPEs are reported relative to the irreducible error $\sigma^2$, with the best possible result being 1.
	
	\begin{table}[ht!]
		\centering
		\caption{MSPE, ${\overline{\text{MSPE}}}$ and $\overline{\text{Cor}}$ of Fast-BSpS and RGLM as a function of the number of models under Scenario 2 with $ \rho=0.5 $ and $\text{SNR} =3 $. \label{tab:MSPE_Groups}} 
		\extrarowsep =2pt
		\begin{tabular}{lrrrrrrrrr}
			\toprule
			& \multicolumn{3}{c}{$\mathbf{\boldsymbol{\zeta}=0.1}$} & \multicolumn{3}{c}{$\mathbf{\boldsymbol{\zeta}=0.2}$} & \multicolumn{3}{c}{$\mathbf{\boldsymbol{\zeta}=0.4}$}\\ \cmidrule(lr){1-1} \cmidrule(lr){2-4} \cmidrule(lr){5-7} \cmidrule(lr){8-10}  \textbf{Method} & \textbf{MSPE} & $\mathbf{\overline{MSPE}}$ & $\mathbf{\overline{Cor}}$ & \textbf{MSPE} & $\mathbf{\overline{MSPE}}$ & $\mathbf{\overline{Cor}}$ & \textbf{MSPE} & $\mathbf{\overline{MSPE}}$ & $\mathbf{\overline{Cor}}$  \\ \cmidrule(lr){1-1} \cmidrule(lr){2-4} \cmidrule(lr){5-7} \cmidrule(lr){8-10} 
			\addlinespace[0.25cm]
			Fast-BSS & 1.39 & $ - $ & $ - $ & 1.30 & $ - $ & $ - $ & 1.24 & $ - $ & $ - $  \\ 
			\addlinespace[0.25cm]
			Fast-BSpS-2 & 1.29 & 1.56 & 0.85 & 1.31 & 1.55 & 0.87 & 1.28 & 1.56 & 0.84 \\ 
			Fast-BSpS-3 & 1.21 & 1.65 & 0.82 & 1.23 & 1.62 & 0.85 & 1.21 & 1.55 & 0.85  \\ 
			Fast-BSpS-4 & 1.23 & 1.77 & 0.80 & 1.20 & 1.70 & 0.83 & 1.19 & 1.65 & 0.83  \\ 
			Fast-BSpS-5 & 1.19 & 1.80 & 0.79 & 1.16 & 1.72 & 0.82 & 1.15 & 1.63 & 0.83  \\ 
			\addlinespace[0.25cm]
			RGLM-2 & 4.34 & 7.37 & 0.25 & 4.38 & 7.51 & 0.27 & 3.50 & 5.95 & 0.34 \\ 
			RGLM-3 & 3.38 & 7.69 & 0.22 & 3.17 & 7.15 & 0.29 & 2.75 & 6.00 & 0.34  \\ 
			RGLM-4 & 2.86 & 7.74 & 0.22 & 2.63 & 6.95 & 0.30 & 2.37 & 6.07 & 0.33  \\ 
			RGLM-5 & 2.47 & 7.50 & 0.23 & 2.30 & 6.69 & 0.31 & 2.12 & 6.13 & 0.34  \\ 
			RGLM-100 & 1.36 & 7.70 & 0.23 & 1.25 & 7.03 & 0.29 & 1.17 & 6.64 & 0.33  \\ 
			\addlinespace[0.25cm]
			\bottomrule
		\end{tabular}
	\end{table}
	
	For Fast-BSpS, it can be seen that for all sparsity levels $\overline{\text{MSPE}}$ increases with the number of models, while MSPE and $\overline{\text{Cor}}$ generally decrease. As the number of models increases, the average accuracy
	of the individual models thus has less impact on the ensemble MSPE compared to $\overline{\text{Cor}}$. This confirms that Fast-BSpS achieves an appropriate balance between individual model accuracy and model diversity, resulting in high accuracy for the ensemble.
	
	For RGLM, the individual models are much weaker, with $\overline{\text{MSPE}}$ being six to seven times the variance of the noise. The individual strength of the models is not controlled for or learned in relation to the number of models in the ensemble. Indeed, the models are equally weak regardless of the number of models. However, due to the random assignment of candidate predictors $\overline{\text{Cor}}$ is much lower for RGLM than for Fast-BSpS, irrespective of the number of models.
	When the number of models becomes large, this average pairwise correlation between the models becomes more important. When $ G = 100 $, RGLM does achieve an ensemble
	MSPE that is lower than the MSPE of Fast-BSS, however it is still
	higher than the ensemble MSPE of Fast-BSpS with only $ G = 2 $ across all sparsity levels. Hence, RGLM  relies on a large number of weak, decorrelated models to achieve a low ensemble MSPE.

	\subsection{Variable Selection}
	
	To elucidate the connections between variable selection and accuracy-diversity tradeoff,  we report in Table \ref{tab:RCPR_Groups} the RC and PR of Fast-BSpS and RGLM as a function of the number of models. It can be seen that Fast-BSpS consistently enjoys very high to perfect PR  and at the same time a high RC relative to RGLM in sparse settings ($\zeta =$ 0.1 or 0.2). For the more dense setting ($\zeta =$ 0.4) PR is lower because the correlation among the active predictors makes it unnecessary to use all of them to achieve a high prediction accuracy.  On the other hand, PR is generally low for RGLM due to the random nature of the methodology. Moreover, RGLM only achieves high RC when many models ($ G=100 $) are used so that all candidate predictors are involved. In this case, RGLM naturally achieves a RC of 1 and PR equal to $\zeta$.
	
	\begin{table}[ht!]
		\centering
		\caption{RC and PR of Fast-BSpS and RGLM as a function of the number of models under Scenario 2 with $ \rho=0.5 $ and $\text{SNR} =3 $. \label{tab:RCPR_Groups}} 
		\extrarowsep =2pt
		\begin{tabular}{lrrrrrr}
			\toprule
			& \multicolumn{2}{c}{$\mathbf{\boldsymbol{\zeta}=0.1}$} & \multicolumn{2}{c}{$\mathbf{\boldsymbol{\zeta}=0.2}$} & \multicolumn{2}{c}{$\mathbf{\boldsymbol{\zeta}=0.4}$}\\ \cmidrule(lr){1-1} \cmidrule(lr){2-3} \cmidrule(lr){4-5} \cmidrule(lr){6-7}   \textbf{Method} & \textbf{RC} & \textbf{PR}  & \textbf{RC} & \textbf{PR}  & \textbf{RC} & \textbf{PR}   \\ \cmidrule(lr){1-1} \cmidrule(lr){2-3} \cmidrule(lr){4-5} \cmidrule(lr){6-7} 
			\addlinespace[0.25cm]
			Fast-BSS & 0.45 & 0.54  & 0.31 & 0.61  & 0.19 & 0.69  \\ 
			\addlinespace[0.25cm]
			Fast-BSpS-2 & 0.56 & 1.00  & 0.28 & 1.00  & 0.16 & 1.00  \\ 
			Fast-BSpS-3 & 0.79 & 0.98 & 0.42 & 1.00  & 0.21 & 1.00   \\ 
			Fast-BSpS-4 & 0.81 & 0.90  & 0.56 & 1.00  & 0.30 & 1.00  \\ 
			Fast-BSpS-5 & 0.84 & 0.85  & 0.67 & 0.99  & 0.34 & 1.00   \\ 
			\addlinespace[0.25cm]
			RGLM-2 & 0.26 & 0.22 & 0.25 & 0.43 & 0.22 & 0.77  \\ 
			RGLM-3 & 0.33 & 0.20 & 0.33 & 0.40 & 0.30 & 0.74   \\ 
			RGLM-4 & 0.40 & 0.19 & 0.42 & 0.40 & 0.39 & 0.76  \\ 
			RGLM-5 & 0.47 & 0.19 & 0.49 & 0.39 & 0.46 & 0.77   \\ 
			RGLM-100 & 1.00 & 0.10 & 1.00 & 0.20 & 1.00 & 0.40  \\ 
			\addlinespace[0.25cm]
			\bottomrule
		\end{tabular}
	\end{table}
	
	Fast-BSpS's ability to achieve high PR and RC in sparse settings reflects its efficient variable selection process, which is directly influenced by its controlled bias, variance, and covariance. By maintaining low bias and variance while ensuring sufficient diversity (lower covariance), Fast-BSpS selects relevant variables more effectively, enhancing both model accuracy and ensemble performance. This means that Fast-BSpS's ensemble model is generally an interpretable model in the sense that most of the variables that appear in the model indeed influence the response. 
	In contrast, RGLM's lower PR and the need for a larger number of models to achieve high RC highlight its reliance on model diversity to compensate for weaker individual models. This underscores how Fast-BSpS's structured approach not only boosts prediction accuracy but also yields good variable selection performance, reinforcing the method's overall efficacy.
	
	\subsection{Computational Cost}
	
	While Fast-BSpS shows very good performance in terms of prediction accuracy, it is expected that it comes at a higher cost due to the challenging $ \ell_0 $-penalized optimization.
	The computational cost (in seconds) of the CV procedure of Fast-BSpS in Scenario 2 of Section \ref{sec:simulation}, across all sparsity levels $\zeta \in \{0.1, 0.2, 0.4\}$, is provided in Table \ref{tab:cpu} as a function of the number of models $G$ in the ensemble. For comparison we include the computational cost of Splitreg-Lasso  which is a multi-convex relaxation of BSpS as explained in Section \ref{sec:multi_convex}  using the \texttt{SplitReg} package \citep{SplitReg} as well as Step-SplitReg as described in Section \ref{sec:stepwise} using the \texttt{stepSplitReg} package. 
	
	The Step-SplitReg method has by far the smallest computational cost and thus is a fast way to generate good initial values for BSpS optimization. The computation time of our \texttt{R} implementation of Fast-BSpS is substantially higher and more sensitive to the number of models than for SplitReg-Lasso. Note that no local combinatorial search is performed in the execution of Algorithm \ref{alg:projected_algo} which would increase the cost further. The computational cost would also increase substantially if a fine grid for the sparsity parameter $ t $ is used. However, our simulation results showed that Fast-BSpS already achieves good performance with the current settings while keeping the computation time reasonable. 
	
	\begin{table}[ht!]
		\centering
		\caption{\label{tab:cpu}Average computation time of \texttt{R} function calls for \texttt{SplitReg}, \texttt{stepSplitReg} and \texttt{PSGD} in CPU seconds for varying number of models. CPU seconds are on a 2.7 GHz Intel Xeon processor in a machine running Linux 7.8 with 125 GB of RAM.} 
		\extrarowsep=2pt
		\begin{tabu}{lrrrrrrr}
			\toprule
			& \multicolumn{4}{c}{Number of Models} \\
			\cmidrule(lr){1-1} \cmidrule(lr){2-5}
			Package     & 2    & 3     & 4     & 5        \\ 
			\cmidrule(lr){1-1} \cmidrule(lr){2-5}
			$ \texttt{SplitReg} $ & 2.23 & 6.56 & 10.41 & 14.91  \\
			$ \texttt{stepSplitReg} $ & 0.25 & 0.69 & 1.05 & 1.17  \\
			$ \texttt{PSGD} $ & 4.67 & 19.38 & 31.43 & 55.92  \\
			\bottomrule
		\end{tabu}
	\end{table}

	\section{Real Data Analysis} \label{sec:eye}
	
	We apply Fast-BSpS and the competitor methods of Section \ref{sec:simulation} on the Bardet-Biedl syndrome (BBS) gene expression dataset \citep{flare}. In \cite{scheetz2006regulation} mutation and functional studies were performed and TRIM32 (tripartite motif-containing protein 32) was identified as a  gene that highly correlates with BBS. Therefore, our purpose is to predict the gene expression level of TRIM32 using the expression levels of $ p = 200 $ genes from mammalian-eye tissue samples identified as relevant by \cite{scheetz2006regulation}.
	
	The full dataset contains $ 120 $ mammalian-eye tissue samples. To fit the models and evaluate their prediction performance, we randomly split the full dataset $ N=50 $ times into a training set of size $ n=30 $ and a test set with the remaining $ m=90 $ samples. For Fast-BSpS we consider the grids $ u\in \{1,2,3,4,5\} $ and $ t\in \{0.3n, 0.4n, 0.5n\}=\{9, 12, 15\} $. For the other methods we use their default settings as in Section \ref{sec:simulation}. We report MSPE for all the methods, averaged over the  $ N=50 $ replications.
	For the ensemble methods we also report $\overline{\text{MSPE}}$ as a measure of individual model accuracy. 
	
\begin{figure}[ht!]
		\centering
		\includegraphics[width=16cm]{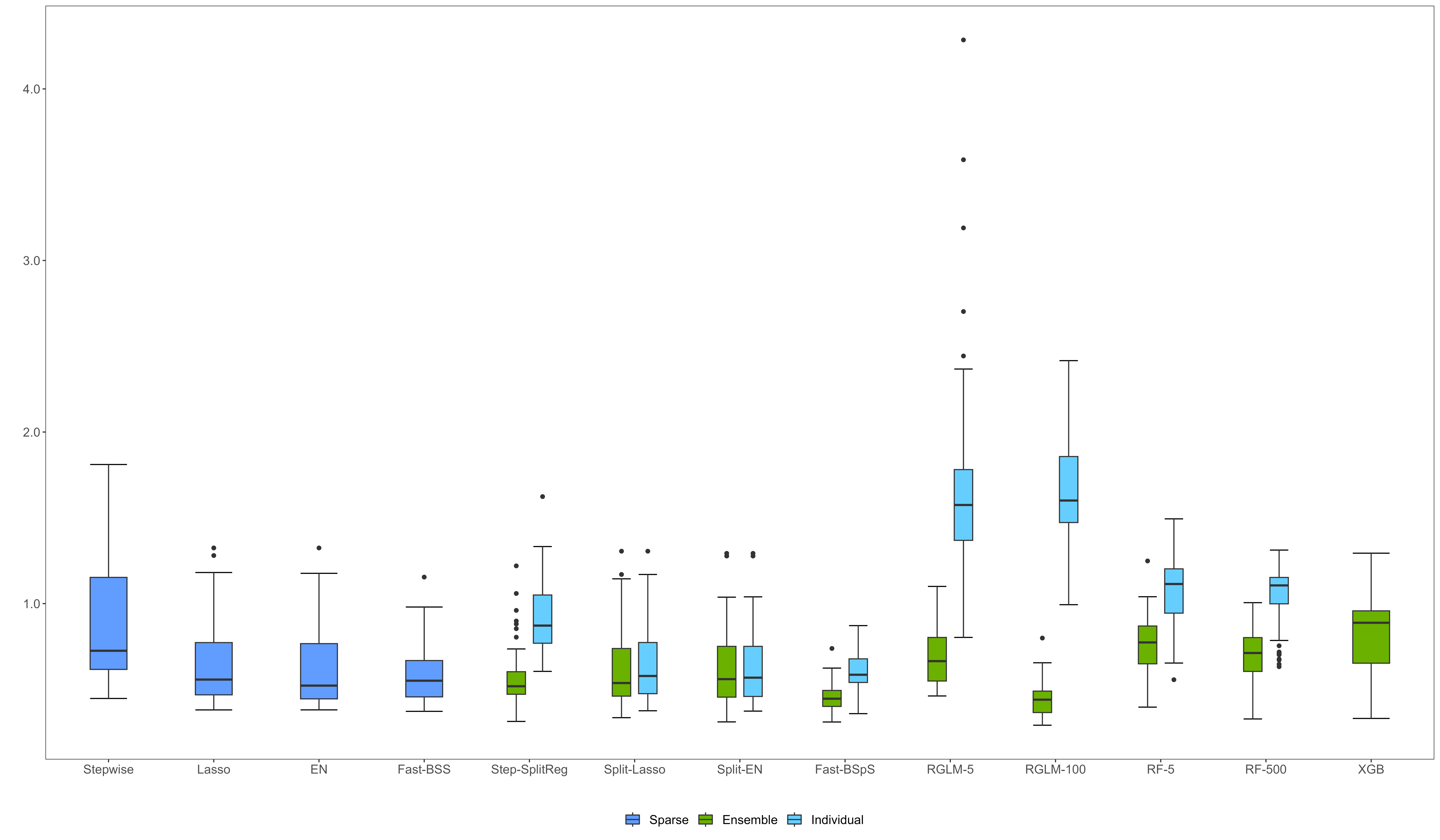}
		\caption{MSPEs of the sparse models, the ensembles and the average of their individual models over $ N=50 $ random splits into training and
			testing sets for the BBS gene expression dataset.}
		\label{fig:BBS_mspe}
	\end{figure}

	The results are reported in Table \ref{tab:trim32_results} and Figure \ref{fig:BBS_mspe}. The two best performances in each column of Table \ref{tab:trim32_results} are marked in bold. 
	Fast-BSpS yields the best MSPE, closely followed by RGLM-100  while all other methods have an MSPE that is at least 10\% higher. 
	While RGLM-100 nearly matches the predictive performance of Fast-BSpS at the ensemble level, its individual model accuracy $\overline{\text{MSPE}}$ is the worst among all methods and is more than double the MSPE of the single model methods in the top part of the table. This is a clear indication that these individual models are not interpretable in the sense that they do not provide insights in the relation between candidate predictors and response, making RGLM a black-box method.
	On the other hand, Fast-BSpS achieves a superior MSPE while at the same time its individual models also achieve a very good accuracy, nearly on par with the accuracy of Fast-BSS the best performing sparse method. In fact, the accuracy of the individual models of Fast-BSpS is  superior to the accuracy of the  Lasso and EN. Fast-BSpS thus not only managed to produce a superior ensemble prediction accuracy with only $ G=5 $ models, but these models are on average as reliable and accurate as standard sparse estimators.
	
	\begin{table}[ht!]
		\centering
		\caption{\label{tab:trim32_results}MSPE and $\overline{\text{MSPE}}$ over the $ N=50 $ random splits into training and testing sets for the BBS gene expression dataset. Standard errors are in parentheses.}
		\extrarowsep =2pt
		\begin{tabular}{lrr}
			\toprule
			\textbf{Method} & \textbf{MSPE} & $\mathbf{\overline{MSPE}}$  \\ 
			\cmidrule(lr){1-1} \cmidrule(lr){2-3} 
			\addlinespace[0.25cm]
			Stepwise & 0.84 (0.30) & $ - $ \\ 
			Lasso & 0.65 (0.25) & $ - $  \\ 
			EN & 0.63 (0.24) & $ - $ \\ 
			Fast-BSS  & 0.59 (0.18) & $ - $  \\ 
			\addlinespace[0.25cm]
			Step-SplitReg & 0.57 (0.19) & 0.92 (0.22)   \\ 
			SplitReg-Lasso & 0.63 (0.24) & 0.65 (0.23)  \\ 
			SplitReg-EN & 0.62 (0.23) & \textbf{0.63 (0.23)}  \\ 
			Fast-BSpS & \textbf{0.45 (0.08)} & \textbf{0.60 (0.10)}  \\ 
			\addlinespace[0.25cm]
			RGLM-5 & 0.69 (0.16) & 1.71 (0.65) \\ 
			RGLM-100 & \textbf{0.45 (0.10)} & 1.67 (0.35) \\ 
			RF-5 & 0.74 (0.19) & 1.04 (0.22) \\ 
			RF-500 & 0.67 (0.17) & 1.03 (0.19) \\ 
			XGB & 0.83 (0.24) & 1.15 (0.23)  \\ 
			\addlinespace[0.25cm]
			\bottomrule
		\end{tabular}
	\end{table}
	
	As seen in Section \ref{sec:empirical_comparison}, the individual models of Fast-BSpS not only have adequate prediction accuracy, but they also tend to select the relevant predictors with high precision. We can exploit this further by using the information from the individual models in Fast-BSpS to rank the genes in order of importance. Consider the sets
	\begin{align} \label{eq:exclusive_sets}
		\mathcal{A}_k = \left\{j:  \sum_{g=1}^{G} \mathbb{I}\left(j \in S^{(g)}\right) \geq k \right\}, \quad 1 \leq k \leq G,
	\end{align}
	Clearly,  $\mathcal{A}_G \subseteq \mathcal{A}_{G-1} \subseteq \cdots \subseteq \mathcal{A}_1$ and $\mathcal{A}_k$ yields the indices of predictors that appear in at least $k$ of the models used by Fast-BSpS. 
	We can now use the sets $\mathcal{A}_k$ to study the distribution of the genes across the different models.  These sets identify genes related to TRIM32 in order of importance because genes appear in more than one model if there are no surrogate genes that may be used to reduce the loss function of BSpS in~(\ref{eq:BSpS}).  Examining the Fast-BSpS solution for the BBS dataset, we find that $ |\mathcal{A}_5| = |\mathcal{A}_4| = 0 $, $|\mathcal{A}_3|=20$, $ |\mathcal{A}_2|=27 $ and $ |\mathcal{A}_1|=28 $. 
	Hence, Fast-BSpS required information from only 28 genes to achieve its excellent performance. Of these 28 genes, 27 appeared in at least two models while 20 genes even appeared in three of the five individual models.	
	Genes shared by more than one model  had the same sign across all the models, re-enforcing the understanding of their relationship with
	the gene expression level of TRIM32. 
	
	To illustrate that Fast-BSpS can identify important genes
	that may be missed by other sparse regression methods, let us consider the 20 genes in $\mathcal{A}_3$ which yield an important
	contribution to the ensemble as they appear in 3 of the individual models. Interestingly, only  4 of these genes appears in the Fast-BSS solution and thus more than half of these 20 important genes would be considered irrelevant for the prediction of the gene expression level of TRIM32 by this method.
	
	\section{Discussion and Future Directions} \label{sec:discussion}
	
	We introduced BSpS, a new methods that builds an ensemble based on a collection of sparse and diverse models learned directly from the data. 
	In the analysis of high-dimensional data, sparse modeling was the main focus in the literature for many years, with many proposals that can be seen as alternative approaches to the NP-hard BSS problem. Our proposal is a generalization of the sparse modeling framework, which simultaneously yields more than one highly interpretable model with good prediction accuracy.	In particular, 
	BSpS can be seen as a generalization of BSS to multiple models that are ensembled. Each of these models  is based on a subset of the candidate predictors in such a way that the sum of their losses is minimized while diversity among the subsets is imposed by a diversity constraint.
	Both the sparsity and the diversity are controlled by an $\ell_0$ constraint, limiting the maximum size of each model
	and the maximum number of models that can share any given predictor, respectively.
	BSpS has the advantage that the ensemble model remains interpretable and furthermore, 
	each individual model presents a potential relationship between a subset of the predictors and the response.
	
	Our ensemble method leverages the principles of the BVC tradeoff to create a well-performing ensemble. By providing control over both model size and predictor usage, our method ensures a balanced adaptive approach to ensemble learning. Our empirical results are in line with theoretical knowledge about ensemble models, demonstrating the effectiveness of our method in achieving low bias, low variance, and low covariance, thereby maximizing overall predictive accuracy. At the same time our method has the advantage of providing interpretable models thanks to its good performance in terms of recall and precision.
	
	Calculating BSpS exactly is an intractable combinatorial problem, thus computational tools to obtain good approximate solutions are needed. Therefore, we adapted a projected subsets gradient descent algorithm to generate approximate solutions to BSpS. The algorithm is well suited for selection of the tuning parameters by CV. Moreover, to generate an initial solution for BSpS in the fully diverse ($ u=1 $) case, we generalized forward stepwise regression to our multiple models setting. 
	Our empirical investigations reveal that our algorithm to calculate approximate BSpS solutions yields ensembles with competitive prediction accuracy and variable selection properties. The data and scripts to replicate the numerical experiments are available at \url{https://doi.org/10.5281/zenodo.13922582}.
	
	
	If problem-specific knowledge is available for certain applications, then this can easily be incorporated into the BSpS method. For example, if certain predictors (e.g. genes) are known to be particularly important or relevant in the prediction of the outcome, this may be easily incorporated by generalizing BSpS in \eqref{eq:BSpS} to 
	\begin{align} \label{eq:BSpS_knowledge}
		\min_{\bbet^1, \dots, \, \bbet^G \in \mathbb{R}^p} \sum_{g=1}^{G} \left\Vert\by - \bX \bbet^g\right\Vert_2^2 \quad \text{subject to} \quad \begin{cases}
			\lVert\bbet^g\rVert_0 \leq t, \, &1 \leq g \leq G, \\
			\lVert\bbet_{j\cdot}\rVert_0 \leq u_j, \, & 1 \leq j \leq p.
		\end{cases}
	\end{align}
	where $ u_j $ is the maximum number of models that may share predictor $ j $, $ 1 \leq j \leq p $. For important genes, the corresponding $u_j$ can then be set equal to $G$ so that they can appear in all models.
	
	The proposed PSGD algorithm is an efficient way to obtain approximate solutions for the nonconvex BSpS problem in \eqref{eq:BSpS}. However, it is still computationally demanding, particularly when the number of groups or the dimension of the data increases. A future area of research is to investigate improvements of the current algorithm, e.g. by adapting the general idea of accelerated proximal gradient descent of \cite{beck2009fast} to projected gradients. 
	Alternative algorithms to split predictors into fully disjoint models may also be explored to generate initial estimators for Fast-BSpS ensembles, see e.g. the split least-angle regression procedure in  \cite{srlars}.
	The BSpS methodology can be extended to a new data-driven ensemble modeling framework that is widely applicable.
	For example, BSpS in \eqref{eq:BSpS} can directly be generalized to GLMs or other parametric settings with some general loss $\mathcal{L}_n$.
	
	\section*{Supplementary Materials}
	
	\hangpara{\parindent}{1}\textbf{Supplementary document:} The supplementary document contains the full results of our simulations.\\
	
	\noindent
	\hangpara{\parindent}{1}\textbf{Materials and Data:} {The data and scripts to replicate the numerical experiments are available at \url{https://doi.org/10.5281/zenodo.13922582}.}	\\
	
	\noindent
	\hangpara{\parindent}{1}\textbf{Code Availability:} The \texttt{R} packages \texttt{stepSplitReg} and \texttt{PSGD} created for this article are
	publicly available on {CRAN} together with their reference manuals at \url{https://CRAN.R-project.org/package=stepSplitReg} and \url{https://CRAN.R-project.org/package=PSGD}, respectively. 
	
	\section*{Funding}
	
	The authors did not receive support from any organization for the submitted work.
	
	\section*{Disclosure Statement}
	
	The authors report there are no competing interests to declare.
	
	\section*{Appendix A: Detailed Results of Simulation Study}
	
	\begin{itemize}
		\item Tables 1-3 contain the detailed simulation results for $ p=500 $, Scenario 1.
		\item Tables 4-6 contain the detailed simulation results for $ p=150 $, Scenario 1.
		\item Tables 7-9 contain the detailed simulation results for $ p=500 $, Scenario 2.
		\item Tables 10-12 contain the detailed simulation results for $ p=150 $, Scenario 2.
	\end{itemize}
	\newpage
	
	\begin{table}[H]
		\centering
		\caption{Mean MSPEs, recalls and precisions for Scenario 1 with $\rho=$ 0.2, $n=$ 50, $p=$ 500. MSPEs maximum standard error is 0.06.} 
		\begin{adjustbox}{totalheight=\textheight-4\baselineskip}
		\resizebox{\textwidth}{!}{
\begin{tabular}{lrrrrrrrrrr}
			\toprule \rule{0pt}{3ex} & &  \multicolumn{3}{c}{$\mathbf{\boldsymbol{\zeta}=0.1}$} & \multicolumn{3}{c}{$\mathbf{\boldsymbol{\zeta}=0.2}$} & \multicolumn{3}{c}{$\mathbf{\boldsymbol{\zeta}=0.4}$} \\ \cmidrule(lr){1-2} \cmidrule(lr){3-5} \cmidrule(lr){6-8} \cmidrule(lr){9-11} $\textbf{SNR}$ & \textbf{Method} & \textbf{MSPE} & \textbf{RC} & \textbf{PR} & \textbf{MSPE} & \textbf{RC} & \textbf{PR} & \textbf{MSPE} & \textbf{RC} & \textbf{PR}  \\ \cmidrule(lr){1-2} \cmidrule(lr){3-5} \cmidrule(lr){6-8} \cmidrule(lr){9-11}  \rule{0pt}{3ex} 
& Stepwise & 1.98 & 0.00 & 0.13 & 1.99 & 0.01 & 0.23 & 1.97 & 0.00 & 0.27 \\ 
   & Fast-BSS & 1.26 & 0.14 & 0.14 & 1.30 & 0.13 & 0.22 & 1.20 & 0.11 & 0.41 \\ 
   & Lasso & 1.36 & 0.07 & 0.15 & 1.35 & 0.07 & 0.25 & 1.33 & 0.06 & 0.42 \\ 
   & Elastic Net & 1.34 & 0.08 & 0.15 & 1.31 & 0.07 & 0.25 & 1.31 & 0.06 & 0.41 \\ 
   & Step-SplitReg & 1.54 & 0.48 & 0.10 & 1.51 & 0.46 & 0.20 & 1.49 & 0.46 & 0.40 \\ 
   & SplitReg-Lasso & 1.25 & 0.19 & 0.14 & 1.21 & 0.18 & 0.23 & 1.24 & 0.17 & 0.42 \\
\bottomrule
		\end{tabular}}
	\end{adjustbox}
	\end{table}
	\begin{table}[H]
		\centering
		\caption{Mean MSPEs, recalls and precisions for Scenario 1 with $\rho=$ 0.5, $n=$ 50, $p=$ 500. MSPEs maximum standard error is 0.07.} 
		\begin{adjustbox}{totalheight=\textheight-4\baselineskip}
		\resizebox{\textwidth}{!}{\begin{tabular}{lrrrrrrrrrr}
			\toprule \rule{0pt}{3ex} & &  \multicolumn{3}{c}{$\mathbf{\boldsymbol{\zeta}=0.1}$} & \multicolumn{3}{c}{$\mathbf{\boldsymbol{\zeta}=0.2}$} & \multicolumn{3}{c}{$\mathbf{\boldsymbol{\zeta}=0.4}$} \\ \cmidrule(lr){1-2} \cmidrule(lr){3-5} \cmidrule(lr){6-8} \cmidrule(lr){9-11} $\textbf{SNR}$ & \textbf{Method} & \textbf{MSPE} & \textbf{RC} & \textbf{PR} & \textbf{MSPE} & \textbf{RC} & \textbf{PR} & \textbf{MSPE} & \textbf{RC} & \textbf{PR}  \\ \cmidrule(lr){1-2} \cmidrule(lr){3-5} \cmidrule(lr){6-8} \cmidrule(lr){9-11}  \rule{0pt}{3ex} 
& Stepwise & 1.54 & 0.01 & 0.16 & 1.61 & 0.01 & 0.25 & 1.61 & 0.00 & 0.36 \\ 
   & Fast-BSS & 1.10 & 0.13 & 0.15 & 1.14 & 0.11 & 0.21 & 1.12 & 0.10 & 0.45 \\ 
   & Lasso & 1.16 & 0.05 & 0.13 & 1.17 & 0.05 & 0.24 & 1.16 & 0.04 & 0.42 \\ 
   & Elastic Net & 1.14 & 0.05 & 0.12 & 1.15 & 0.05 & 0.22 & 1.15 & 0.05 & 0.42 \\ 
   & Step-SplitReg & 1.42 & 0.45 & 0.10 & 1.43 & 0.46 & 0.20 & 1.44 & 0.45 & 0.41 \\ 
   & SplitReg-Lasso & 1.10 & 0.12 & 0.12 & 1.11 & 0.13 & 0.23 & 1.10 & 0.12 & 0.42 \\
\bottomrule
		\end{tabular}}
	\end{adjustbox}
	\end{table}
	\begin{table}[H]
		\centering
		\caption{Mean MSPEs, recalls and precisions for Scenario 1 with $\rho=$ 0.8, $n=$ 50, $p=$ 500. MSPEs maximum standard error is 0.11.} 
		\begin{adjustbox}{totalheight=\textheight-4\baselineskip}
		\resizebox{\textwidth}{!}{\begin{tabular}{lrrrrrrrrrr}
			\toprule \rule{0pt}{3ex} & &  \multicolumn{3}{c}{$\mathbf{\boldsymbol{\zeta}=0.1}$} & \multicolumn{3}{c}{$\mathbf{\boldsymbol{\zeta}=0.2}$} & \multicolumn{3}{c}{$\mathbf{\boldsymbol{\zeta}=0.4}$} \\ \cmidrule(lr){1-2} \cmidrule(lr){3-5} \cmidrule(lr){6-8} \cmidrule(lr){9-11} $\textbf{SNR}$ & \textbf{Method} & \textbf{MSPE} & \textbf{RC} & \textbf{PR} & \textbf{MSPE} & \textbf{RC} & \textbf{PR} & \textbf{MSPE} & \textbf{RC} & \textbf{PR}  \\ \cmidrule(lr){1-2} \cmidrule(lr){3-5} \cmidrule(lr){6-8} \cmidrule(lr){9-11}  \rule{0pt}{3ex} 
& Stepwise & 1.22 & 0.00 & 0.15 & 1.26 & 0.00 & 0.22 & 1.28 & 0.00 & 0.47 \\ 
   & Fast-BSS & 1.11 & 0.10 & 0.10 & 1.09 & 0.07 & 0.20 & 1.10 & 0.09 & 0.43 \\ 
   & Lasso & 1.08 & 0.04 & 0.14 & 1.10 & 0.03 & 0.22 & 1.09 & 0.03 & 0.43 \\ 
   & Elastic Net & 1.06 & 0.04 & 0.12 & 1.09 & 0.04 & 0.22 & 1.09 & 0.04 & 0.43 \\ 
   & Step-SplitReg & 1.34 & 0.43 & 0.10 & 1.34 & 0.40 & 0.20 & 1.40 & 0.41 & 0.40 \\ 
   & SplitReg-Lasso & 1.03 & 0.09 & 0.12 & 1.06 & 0.10 & 0.22 & 1.06 & 0.08 & 0.44 \\
\bottomrule
		\end{tabular}}
	\end{adjustbox}
	\end{table}

	\begin{table}[H]
		\centering
		\caption{Mean MSPEs, recalls and precisions for Scenario 1 with $\rho=$ 0.2, $n=$ 50, $p=$ 150. MSPEs maximum standard error is 0.07.} 
		\begin{adjustbox}{totalheight=\textheight-4\baselineskip}
		\resizebox{\textwidth}{!}{\begin{tabular}{lrrrrrrrrrr}
			\toprule \rule{0pt}{3ex} & &  \multicolumn{3}{c}{$\mathbf{\boldsymbol{\zeta}=0.1}$} & \multicolumn{3}{c}{$\mathbf{\boldsymbol{\zeta}=0.2}$} & \multicolumn{3}{c}{$\mathbf{\boldsymbol{\zeta}=0.4}$} \\ \cmidrule(lr){1-2} \cmidrule(lr){3-5} \cmidrule(lr){6-8} \cmidrule(lr){9-11} $\textbf{SNR}$ & \textbf{Method} & \textbf{MSPE} & \textbf{RC} & \textbf{PR} & \textbf{MSPE} & \textbf{RC} & \textbf{PR} & \textbf{MSPE} & \textbf{RC} & \textbf{PR}  \\ \cmidrule(lr){1-2} \cmidrule(lr){3-5} \cmidrule(lr){6-8} \cmidrule(lr){9-11}  \rule{0pt}{3ex} 
& Stepwise & 2.17 & 0.07 & 0.44 & 2.00 & 0.02 & 0.22 & 1.91 & 0.02 & 0.34 \\ 
   & Fast-BSS & 1.77 & 0.34 & 0.29 & 1.43 & 0.36 & 0.25 & 1.26 & 0.39 & 0.46 \\ 
   & Lasso & 1.74 & 0.26 & 0.25 & 1.55 & 0.16 & 0.28 & 1.33 & 0.15 & 0.46 \\ 
   & Elastic Net & 1.77 & 0.29 & 0.24 & 1.50 & 0.17 & 0.28 & 1.33 & 0.17 & 0.46 \\ 
   & Step-SplitReg & 1.73 & 0.53 & 0.15 & 1.51 & 0.41 & 0.23 & 1.34 & 0.42 & 0.41 \\ 
   & SplitReg-Lasso & 1.70 & 0.37 & 0.23 & 1.43 & 0.35 & 0.30 & 1.22 & 0.35 & 0.46 \\
\bottomrule
		\end{tabular}}
	\end{adjustbox}
	\end{table}
	\begin{table}[H]
		\centering
		\caption{Mean MSPEs, recalls and precisions for Scenario 1 with $\rho=$ 0.5, $n=$ 50, $p=$ 150. MSPEs maximum standard error is 0.12.} 
		\begin{adjustbox}{totalheight=\textheight-4\baselineskip}
		\resizebox{\textwidth}{!}{\begin{tabular}{lrrrrrrrrrr}
			\toprule \rule{0pt}{3ex} & &  \multicolumn{3}{c}{$\mathbf{\boldsymbol{\zeta}=0.1}$} & \multicolumn{3}{c}{$\mathbf{\boldsymbol{\zeta}=0.2}$} & \multicolumn{3}{c}{$\mathbf{\boldsymbol{\zeta}=0.4}$} \\ \cmidrule(lr){1-2} \cmidrule(lr){3-5} \cmidrule(lr){6-8} \cmidrule(lr){9-11} $\textbf{SNR}$ & \textbf{Method} & \textbf{MSPE} & \textbf{RC} & \textbf{PR} & \textbf{MSPE} & \textbf{RC} & \textbf{PR} & \textbf{MSPE} & \textbf{RC} & \textbf{PR}  \\ \cmidrule(lr){1-2} \cmidrule(lr){3-5} \cmidrule(lr){6-8} \cmidrule(lr){9-11}  \rule{0pt}{3ex} 
			& Stepwise & 1.72 & 0.05 & 0.34 & 1.57 & 0.02 & 0.19 & 1.50 & 0.02 & 0.43 \\ 
   & Fast-BSS & 1.43 & 0.30 & 0.21 & 1.19 & 0.29 & 0.22 & 1.12 & 0.28 & 0.48 \\ 
   & Lasso & 1.42 & 0.18 & 0.19 & 1.21 & 0.12 & 0.24 & 1.14 & 0.11 & 0.43 \\ 
   & Elastic Net & 1.46 & 0.21 & 0.19 & 1.19 & 0.12 & 0.24 & 1.17 & 0.13 & 0.43 \\ 
   & Step-SplitReg & 1.46 & 0.42 & 0.14 & 1.25 & 0.34 & 0.23 & 1.21 & 0.34 & 0.42 \\ 
   & SplitReg-Lasso & 1.39 & 0.33 & 0.16 & 1.14 & 0.30 & 0.25 & 1.08 & 0.26 & 0.45 \\
\bottomrule
		\end{tabular}}
	\end{adjustbox}
	\end{table}
	\begin{table}[H]
		\centering
		\caption{Mean MSPEs, recalls and precisions for Scenario 1 with $\rho=$ 0.8, $n=$ 50, $p=$ 150. MSPEs maximum standard error is 0.09.} 
		\begin{adjustbox}{totalheight=\textheight-4\baselineskip}
		\resizebox{\textwidth}{!}{\begin{tabular}{lrrrrrrrrrr}
			\toprule \rule{0pt}{3ex} & &  \multicolumn{3}{c}{$\mathbf{\boldsymbol{\zeta}=0.1}$} & \multicolumn{3}{c}{$\mathbf{\boldsymbol{\zeta}=0.2}$} & \multicolumn{3}{c}{$\mathbf{\boldsymbol{\zeta}=0.4}$} \\ \cmidrule(lr){1-2} \cmidrule(lr){3-5} \cmidrule(lr){6-8} \cmidrule(lr){9-11} $\textbf{SNR}$ & \textbf{Method} & \textbf{MSPE} & \textbf{RC} & \textbf{PR} & \textbf{MSPE} & \textbf{RC} & \textbf{PR} & \textbf{MSPE} & \textbf{RC} & \textbf{PR}  \\ \cmidrule(lr){1-2} \cmidrule(lr){3-5} \cmidrule(lr){6-8} \cmidrule(lr){9-11}  \rule{0pt}{3ex} & Stepwise & 1.37 & 0.02 & 0.21 & 1.24 & 0.01 & 0.19 & 1.24 & 0.01 & 0.40 \\ 
   & Fast-BSS & 1.22 & 0.29 & 0.17 & 1.05 & 0.24 & 0.22 & 1.08 & 0.26 & 0.44 \\ 
   & Lasso & 1.26 & 0.13 & 0.18 & 1.08 & 0.08 & 0.21 & 1.07 & 0.08 & 0.42 \\ 
   & Elastic Net & 1.27 & 0.14 & 0.17 & 1.06 & 0.08 & 0.22 & 1.06 & 0.10 & 0.43 \\ 
   & Step-SplitReg & 1.28 & 0.30 & 0.14 & 1.14 & 0.28 & 0.22 & 1.13 & 0.26 & 0.42 \\ 
   & SplitReg-Lasso & 1.22 & 0.23 & 0.13 & 1.05 & 0.19 & 0.23 & 1.04 & 0.22 & 0.44 \\
\bottomrule
		\end{tabular}}
	\end{adjustbox}
	\end{table}

	\begin{table}[H]
		\centering
		\caption{Mean MSPEs, recalls and precisions for Scenario 2 with $\rho=$ 0.2, $n=$ 50, $p=$ 500. MSPEs maximum standard error is 0.08.} 
		\begin{adjustbox}{totalheight=\textheight-4\baselineskip}
			\resizebox{\textwidth}{!}{\begin{tabular}{lrrrrrrrrrr}
					\toprule \rule{0pt}{3ex} & &  \multicolumn{3}{c}{$\mathbf{\boldsymbol{\zeta}=0.1}$} & \multicolumn{3}{c}{$\mathbf{\boldsymbol{\zeta}=0.2}$} & \multicolumn{3}{c}{$\mathbf{\boldsymbol{\zeta}=0.4}$} \\ \cmidrule(lr){1-2} \cmidrule(lr){3-5} \cmidrule(lr){6-8} \cmidrule(lr){9-11} $\textbf{SNR}$ & \textbf{Method} & \textbf{MSPE} & \textbf{RC} & \textbf{PR} & \textbf{MSPE} & \textbf{RC} & \textbf{PR} & \textbf{MSPE} & \textbf{RC} & \textbf{PR}  \\ \cmidrule(lr){1-2} \cmidrule(lr){3-5} \cmidrule(lr){6-8} \cmidrule(lr){9-11}  \rule{0pt}{3ex} 
					& Stepwise & 1.97 & 0.01 & 0.51 & 2.00 & 0.01 & 0.46 & 2.00 & 0.01 & 0.55 \\ 
   & Fast-BSS & 1.49 & 0.31 & 0.39 & 1.37 & 0.25 & 0.46 & 1.29 & 0.16 & 0.59 \\ 
   & Lasso & 1.54 & 0.17 & 0.49 & 1.44 & 0.14 & 0.61 & 1.41 & 0.09 & 0.70 \\ 
   & Elastic Net & 1.52 & 0.19 & 0.48 & 1.41 & 0.15 & 0.63 & 1.39 & 0.09 & 0.71 \\ 
   & Step-SplitReg & 1.67 & 0.54 & 0.12 & 1.57 & 0.50 & 0.22 & 1.52 & 0.45 & 0.40 \\ 
   & SplitReg-Lasso & 1.49 & 0.29 & 0.48 & 1.32 & 0.34 & 0.53 & 1.28 & 0.23 & 0.68 \\
\bottomrule
			\end{tabular}}
		\end{adjustbox}
	\end{table}
	\begin{table}[H]
		\centering
		\caption{Mean MSPEs, recalls and precisions for Scenario 2 with $\rho=$ 0.5, $n=$ 50, $p=$ 500. MSPEs maximum standard error is 0.07.} 
		\begin{adjustbox}{totalheight=\textheight-4\baselineskip}
			\resizebox{\textwidth}{!}{\begin{tabular}{lrrrrrrrrrr}
					\toprule \rule{0pt}{3ex} & &  \multicolumn{3}{c}{$\mathbf{\boldsymbol{\zeta}=0.1}$} & \multicolumn{3}{c}{$\mathbf{\boldsymbol{\zeta}=0.2}$} & \multicolumn{3}{c}{$\mathbf{\boldsymbol{\zeta}=0.4}$} \\ \cmidrule(lr){1-2} \cmidrule(lr){3-5} \cmidrule(lr){6-8} \cmidrule(lr){9-11} $\textbf{SNR}$ & \textbf{Method} & \textbf{MSPE} & \textbf{RC} & \textbf{PR} & \textbf{MSPE} & \textbf{RC} & \textbf{PR} & \textbf{MSPE} & \textbf{RC} & \textbf{PR}  \\ \cmidrule(lr){1-2} \cmidrule(lr){3-5} \cmidrule(lr){6-8} \cmidrule(lr){9-11}  \rule{0pt}{3ex} 
					& Stepwise & 1.64 & 0.02 & 0.80 & 1.64 & 0.01 & 0.81 & 1.77 & 0.01 & 0.60 \\ 
   & Fast-BSS & 1.25 & 0.29 & 0.51 & 1.20 & 0.24 & 0.53 & 1.19 & 0.12 & 0.63 \\ 
   & Lasso & 1.24 & 0.16 & 0.61 & 1.25 & 0.10 & 0.62 & 1.23 & 0.06 & 0.69 \\ 
   & Elastic Net & 1.23 & 0.18 & 0.59 & 1.21 & 0.12 & 0.68 & 1.21 & 0.07 & 0.69 \\ 
   & Step-SplitReg & 1.43 & 0.51 & 0.11 & 1.45 & 0.45 & 0.20 & 1.44 & 0.41 & 0.38 \\ 
   & SplitReg-Lasso & 1.21 & 0.33 & 0.58 & 1.16 & 0.26 & 0.61 & 1.17 & 0.15 & 0.72 \\
\bottomrule
			\end{tabular}}
		\end{adjustbox}
	\end{table}
	\begin{table}[H]
		\centering
		\caption{Mean MSPEs, recalls and precisions for Scenario 2 with $\rho=$ 0.8, $n=$ 50, $p=$ 500. MSPEs maximum standard error is 0.05.} 
		\begin{adjustbox}{totalheight=\textheight-4\baselineskip}
			\resizebox{\textwidth}{!}{\begin{tabular}{lrrrrrrrrrr}
					\toprule \rule{0pt}{3ex} & &  \multicolumn{3}{c}{$\mathbf{\boldsymbol{\zeta}=0.1}$} & \multicolumn{3}{c}{$\mathbf{\boldsymbol{\zeta}=0.2}$} & \multicolumn{3}{c}{$\mathbf{\boldsymbol{\zeta}=0.4}$} \\ \cmidrule(lr){1-2} \cmidrule(lr){3-5} \cmidrule(lr){6-8} \cmidrule(lr){9-11} $\textbf{SNR}$ & \textbf{Method} & \textbf{MSPE} & \textbf{RC} & \textbf{PR} & \textbf{MSPE} & \textbf{RC} & \textbf{PR} & \textbf{MSPE} & \textbf{RC} & \textbf{PR}  \\ \cmidrule(lr){1-2} \cmidrule(lr){3-5} \cmidrule(lr){6-8} \cmidrule(lr){9-11}  \rule{0pt}{3ex} 
					& Stepwise & 1.24 & 0.02 & 0.93 & 1.27 & 0.01 & 0.91 & 1.33 & 0.00 & 0.85 \\ 
   & Fast-BSS & 1.12 & 0.25 & 0.64 & 1.16 & 0.19 & 0.63 & 1.12 & 0.11 & 0.67 \\ 
   & Lasso & 1.14 & 0.11 & 0.58 & 1.18 & 0.06 & 0.57 & 1.18 & 0.03 & 0.61 \\ 
   & Elastic Net & 1.12 & 0.14 & 0.59 & 1.20 & 0.09 & 0.59 & 1.17 & 0.05 & 0.63 \\ 
   & Step-SplitReg & 1.39 & 0.38 & 0.09 & 1.41 & 0.38 & 0.17 & 1.42 & 0.38 & 0.35 \\ 
   & SplitReg-Lasso & 1.11 & 0.27 & 0.59 & 1.13 & 0.15 & 0.61 & 1.15 & 0.09 & 0.66 \\
\bottomrule
			\end{tabular}}
		\end{adjustbox}
	\end{table}

	\begin{table}[H]
		\centering
		\caption{Mean MSPEs, recalls and precisions for Scenario 2 with $\rho=$ 0.2, $n=$ 50, $p=$ 150. MSPEs maximum standard error is 0.09.} 
		\begin{adjustbox}{totalheight=\textheight-4\baselineskip}
			\resizebox{\textwidth}{!}{\begin{tabular}{lrrrrrrrrrr}
					\toprule \rule{0pt}{3ex} & &  \multicolumn{3}{c}{$\mathbf{\boldsymbol{\zeta}=0.1}$} & \multicolumn{3}{c}{$\mathbf{\boldsymbol{\zeta}=0.2}$} & \multicolumn{3}{c}{$\mathbf{\boldsymbol{\zeta}=0.4}$} \\ \cmidrule(lr){1-2} \cmidrule(lr){3-5} \cmidrule(lr){6-8} \cmidrule(lr){9-11} $\textbf{SNR}$ & \textbf{Method} & \textbf{MSPE} & \textbf{RC} & \textbf{PR} & \textbf{MSPE} & \textbf{RC} & \textbf{PR} & \textbf{MSPE} & \textbf{RC} & \textbf{PR}  \\ \cmidrule(lr){1-2} \cmidrule(lr){3-5} \cmidrule(lr){6-8} \cmidrule(lr){9-11}  \rule{0pt}{3ex} 
					& Stepwise & 2.17 & 0.08 & 0.56 & 1.96 & 0.03 & 0.48 & 1.95 & 0.03 & 0.59 \\ 
   & Fast-BSS & 1.83 & 0.38 & 0.52 & 1.53 & 0.47 & 0.41 & 1.31 & 0.45 & 0.57 \\ 
   & Lasso & 1.78 & 0.36 & 0.46 & 1.62 & 0.25 & 0.52 & 1.43 & 0.21 & 0.65 \\ 
   & Elastic Net & 1.75 & 0.40 & 0.44 & 1.58 & 0.27 & 0.56 & 1.41 & 0.23 & 0.63 \\ 
   & Step-SplitReg & 1.80 & 0.67 & 0.17 & 1.56 & 0.57 & 0.27 & 1.36 & 0.48 & 0.46 \\ 
   & SplitReg-Lasso & 1.74 & 0.45 & 0.46 & 1.56 & 0.37 & 0.54 & 1.34 & 0.41 & 0.63 \\
\bottomrule
			\end{tabular}}
		\end{adjustbox}
	\end{table}
	\begin{table}[H]
		\centering
		\caption{Mean MSPEs, recalls and precisions for Scenario 2 with $\rho=$ 0.5, $n=$ 50, $p=$ 150. MSPEs maximum standard error is 0.08.} 
		\begin{adjustbox}{totalheight=\textheight-4\baselineskip}
			\resizebox{\textwidth}{!}{\begin{tabular}{lrrrrrrrrrr}
					\toprule \rule{0pt}{3ex} & &  \multicolumn{3}{c}{$\mathbf{\boldsymbol{\zeta}=0.1}$} & \multicolumn{3}{c}{$\mathbf{\boldsymbol{\zeta}=0.2}$} & \multicolumn{3}{c}{$\mathbf{\boldsymbol{\zeta}=0.4}$} \\ \cmidrule(lr){1-2} \cmidrule(lr){3-5} \cmidrule(lr){6-8} \cmidrule(lr){9-11} $\textbf{SNR}$ & \textbf{Method} & \textbf{MSPE} & \textbf{RC} & \textbf{PR} & \textbf{MSPE} & \textbf{RC} & \textbf{PR} & \textbf{MSPE} & \textbf{RC} & \textbf{PR}  \\ \cmidrule(lr){1-2} \cmidrule(lr){3-5} \cmidrule(lr){6-8} \cmidrule(lr){9-11}  \rule{0pt}{3ex} 
					& Stepwise & 1.76 & 0.07 & 0.78 & 1.73 & 0.03 & 0.67 & 1.67 & 0.02 & 0.68 \\ 
   & Fast-BSS & 1.52 & 0.37 & 0.60 & 1.29 & 0.46 & 0.47 & 1.17 & 0.39 & 0.59 \\ 
   & Lasso & 1.45 & 0.34 & 0.54 & 1.28 & 0.23 & 0.65 & 1.22 & 0.16 & 0.67 \\ 
   & Elastic Net & 1.47 & 0.38 & 0.48 & 1.29 & 0.26 & 0.64 & 1.24 & 0.18 & 0.64 \\ 
   & Step-SplitReg & 1.51 & 0.66 & 0.18 & 1.33 & 0.50 & 0.27 & 1.24 & 0.39 & 0.44 \\ 
   & SplitReg-Lasso & 1.45 & 0.41 & 0.55 & 1.25 & 0.43 & 0.62 & 1.14 & 0.42 & 0.70 \\
\bottomrule
			\end{tabular}}
		\end{adjustbox}
	\end{table}
	\begin{table}[H]
		\centering
		\caption{Mean MSPEs, recalls and precisions for Scenario 2 with $\rho=$ 0.8, $n=$ 50, $p=$ 150. MSPEs maximum standard error is 0.07.} 
		\begin{adjustbox}{totalheight=\textheight-4\baselineskip}
			\resizebox{\textwidth}{!}{\begin{tabular}{lrrrrrrrrrr}
					\toprule \rule{0pt}{3ex} & &  \multicolumn{3}{c}{$\mathbf{\boldsymbol{\zeta}=0.1}$} & \multicolumn{3}{c}{$\mathbf{\boldsymbol{\zeta}=0.2}$} & \multicolumn{3}{c}{$\mathbf{\boldsymbol{\zeta}=0.4}$} \\ \cmidrule(lr){1-2} \cmidrule(lr){3-5} \cmidrule(lr){6-8} \cmidrule(lr){9-11} $\textbf{SNR}$ & \textbf{Method} & \textbf{MSPE} & \textbf{RC} & \textbf{PR} & \textbf{MSPE} & \textbf{RC} & \textbf{PR} & \textbf{MSPE} & \textbf{RC} & \textbf{PR}  \\ \cmidrule(lr){1-2} \cmidrule(lr){3-5} \cmidrule(lr){6-8} \cmidrule(lr){9-11}  \rule{0pt}{3ex} 
					& Stepwise & 1.34 & 0.07 & 0.93 & 1.26 & 0.03 & 0.86 & 1.22 & 0.02 & 0.90 \\ 
   & Fast-BSS & 1.41 & 0.34 & 0.58 & 1.18 & 0.46 & 0.51 & 1.11 & 0.31 & 0.60 \\ 
   & Lasso & 1.35 & 0.26 & 0.53 & 1.19 & 0.16 & 0.58 & 1.16 & 0.10 & 0.59 \\ 
   & Elastic Net & 1.33 & 0.33 & 0.52 & 1.18 & 0.20 & 0.61 & 1.15 & 0.12 & 0.61 \\ 
   & Step-SplitReg & 1.37 & 0.51 & 0.16 & 1.22 & 0.37 & 0.23 & 1.16 & 0.28 & 0.40 \\ 
   & SplitReg-Lasso & 1.32 & 0.43 & 0.53 & 1.17 & 0.35 & 0.62 & 1.11 & 0.28 & 0.65 \\
\bottomrule
			\end{tabular}}
		\end{adjustbox}
	\end{table}

	\bibliographystyle{Chicago}
	\bibliography{Multi_Model_Subset_Selection}
	
\end{document}